\documentclass[twocolumn,aps,prl,superscriptaddress,groupedaddress]{revtex4}
\usepackage{dcolumn}
\usepackage{bm}
\usepackage{amsmath,amssymb}
\usepackage[dvipdfmx]{graphicx}
\usepackage{color}
\usepackage{soul}
\sethlcolor{yellow}

\begin{document}

\title{Three-body correlations in nonlinear response of correlated quantum liquid}
\author{Tokuro Hata\footnote{Correspondence and requests for materials should be addressed to T.H. (email: hata@phys.titech.ac.jp) or K.K. (email: kensuke@phys.s.u-tokyo.ac.jp
)}}
\affiliation{Graduate School of Science, Osaka University, Toyonaka, Osaka 560-0043, Japan.}

\author{Yoshimichi Teratani}
\affiliation{Department of Physics, Osaka City University, Osaka 558-8585, Japan.}

\author{Tomonori Arakawa}
\affiliation{Graduate School of Science, Osaka University, Toyonaka, Osaka 560-0043, Japan.}
\affiliation{Center for Spintronics Research Network, Osaka University, Toyonaka, Osaka 560-8531, Japan.}

\author{Sanghyun Lee}
\affiliation{Graduate School of Science, Osaka University, Toyonaka, Osaka 560-0043, Japan.}

\author{Meydi Ferrier}
\affiliation{Graduate School of Science, Osaka University, Toyonaka, Osaka 560-0043, Japan.}
\affiliation{Université Paris-Saclay, CNRS, Laboratoire de Physique des Solides, 91405, Orsay, France.}

\author{Richard Deblock}
\affiliation{Université Paris-Saclay, CNRS, Laboratoire de Physique des Solides, 91405, Orsay, France.}

\author{Rui Sakano}
\affiliation{The Institute for Solid State Physics, The University of Tokyo, Chiba 277-8581, Japan.}

\author{Akira Oguri}
\affiliation{Department of Physics, Osaka City University, Osaka 558-8585, Japan.}
\affiliation{Nambu Yoichiro Institute of Theoretical and Experimental Physics, Osaka City University, Osaka 558-8585, Japan.}

\author{Kensuke Kobayashi}
\affiliation{Graduate School of Science, Osaka University, Toyonaka, Osaka 560-0043, Japan.}
\affiliation{Institute for Physics of Intelligence and Department of Physics, The University of Tokyo, Bunkyo-ku, Tokyo 113-0033, Japan.}
\affiliation{Trans-scale Quantum Science Institute, The University of Tokyo, Bunkyo-ku, Tokyo 113-0033, Japan.}

\date{\today}

\begin{abstract}
\subsection*{Abstract}
Behavior of quantum liquids is a fascinating topic in physics. Even in a strongly correlated case, the linear response of a given system to an external field is described by the fluctuation-dissipation relations based on the two-body correlations in the equilibrium. However, to explore nonlinear non-equilibrium behaviors of the system beyond this well-established regime, the role of higher order correlations starting from the three-body correlations must be revealed. In this work, we experimentally investigate a controllable quantum liquid realized in a Kondo-correlated quantum dot and prove the relevance of the three-body correlations in the nonlinear conductance at finite magnetic field, which validates the recent Fermi liquid theory extended to the non-equilibrium regime.
\end{abstract}

\maketitle
\subsection*{{\large {Introduction}}}
Understanding the properties of correlated quantum liquids is a fundamental issue of condensed matter physics. Even in a strongly correlated case, fascinatingly, we can tell that the equilibrium fluctuations of the system govern its linear response to an external field, relying on the fluctuation-dissipation relations based on the two-body correlations. Going beyond this well-established regime, the three-body correlations are known to be of importance for van der Waals force~\cite{Axilrod1943}, the three-body force in nuclei~\cite{Otsuka2010}, the Efimov state~\cite{Efimov, Kraemer2006}, the ring exchange interaction in solid $^3$He~\cite{Thouless_1965, Roger1983}, and frustrated spin systems~\cite{Lacroix2011}.

A quantum dot (QD) shows the Kondo effect, when a localized spin in the QD is coupled with conduction electrons in the reservoirs to form a spin singlet state~\cite{GoldhaberGordonNature1998,Cronenwett540,SCHMID1998182,vanderWielScience2000}. The QD in the Kondo regime is an ideal realization of a strongly correlated quantum liquid with high controllability and accessibility to the spin degree of freedom.
The essential physics can be captured by the Anderson impurity model, which describes the single impurity state with spin $\sigma~(=\uparrow, \downarrow$ or $\pm1)$ and energy $\varepsilon_\sigma$ coupled to the two metallic reservoirs.
The Fermi liquid theory for the Anderson model, which has been developed phenomenologically~\cite{NozieresJLTP1974} and  microscopically~\cite{YosidaPTPS1970,YamadaPTP1975_1,YamadaPTP1975_2, ShibaPTP1975, YoshimoriPTP1976}, can explain the low-energy physics of this highly correlated quantum system as the properties of an ensemble of quasi-particles interacting with each other via residual interaction.

\begin{figure*}[t]
\center \includegraphics[width=0.95\linewidth]{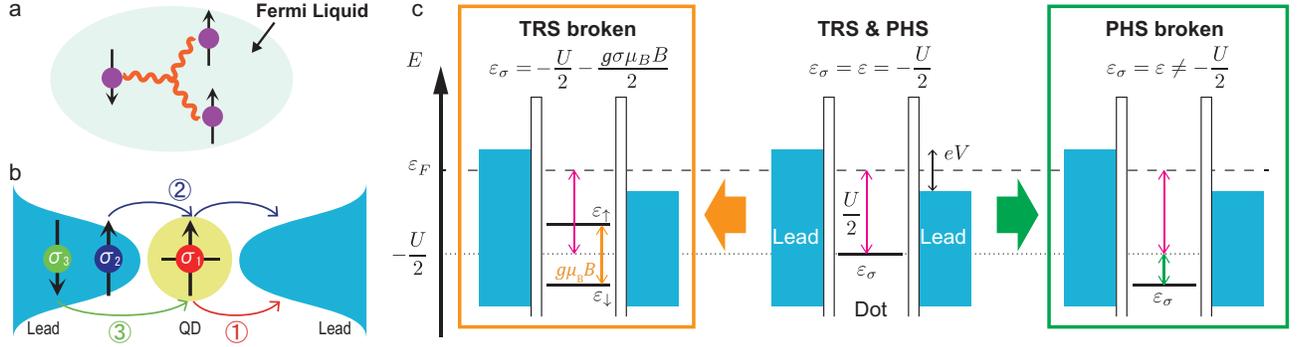}
\caption{{\bf {Three-body correlations and symmetry breaking in a quantum dot in the Kondo regime.} }{\bf a} {Schematic view of the correlations between the three electrons accounting for the Fermi liquid correction.} {\bf b} Schematic view of the three-body correlation $\chi_{\uparrow\uparrow\downarrow}$. Electrons fluctuate between the quantum dot (QD) and the leads in the equilibrium and all those processes are summed up in  $\chi_{\uparrow\uparrow\downarrow}$. {\bf c} (center) QD with the time-reversal symmetry (TRS) and particle-hole symmetry (PHS). The energy level is $\varepsilon_{\sigma}=-U/2$. The two-body correlations $\chi_{\sigma_1\sigma_2}$ are finite, while the three-body correlations are quenched. (left) QD with broken TRS, where the two spins are separated by $g\mu_{\rm B} B$. (right) QD with broken PHS, where the energy level is $\varepsilon \neq-U/2$. Experimentally, the gate voltage is applied to the QD in order to tune the energy level. In these broken symmetry cases, the three-body correlations $\chi_{\sigma_1\sigma_2\sigma_3}$ are finite, which we detect in this paper. }
\label{001}
\end{figure*}

At zero magnetic field, namely when the time-reversal symmetry (TRS) is present, the linear response of the Kondo state can be explained in terms of only a few parameters; the phase shift $\delta_{\sigma}$ of the electron passing through the impurity state, and the two-body correlations ($\chi_{\sigma_1\sigma_2}$) between spin $\sigma_1$ and spin $\sigma_2$.
Here, the two-body correlation is identical to the conventional linear susceptibility, which is defined as the differential of the free energy $\Omega$ of the total system regarding the energies of two electrons, $\chi_{\sigma_1 \sigma_2} \equiv - \partial^2 \Omega/\partial \varepsilon_{\sigma_1}\partial \varepsilon_{\sigma_2}$ (see Supplementary Note 1).
These parameters determine the two crucial physical quantities of the Kondo problem:  the Kondo temperature $k_{\rm B}T_{\rm K} \equiv 1/(4\chi_{\uparrow\uparrow})$ and the Wilson ratio $R\equiv 1-\chi_{\uparrow\downarrow}/\chi_{\uparrow\uparrow}$~\cite{NozieresJLTP1974, YosidaPTPS1970}.

To go beyond this well-established regime, nonlinear susceptibility, that is, three-body correlation $\chi_{\sigma_1\sigma_2\sigma_3}$ comes into the game (see Fig.~\ref{001}a).
Similarly as above, it is defined as $\chi_{\sigma_1 \sigma_2 \sigma_3} \equiv - \partial^3 \Omega/\partial \varepsilon_{\sigma_1}\partial \varepsilon_{\sigma_2}\partial \varepsilon_{\sigma_3}$.
To give an intuitive picture, consider a situation where three electrons with spin $\sigma_1$, $\sigma_2$, and $\sigma_3$ pass through the QD level in sequence as schematically shown in Fig.~\ref{001}b.
The electron with $\sigma_1$ first occupies and leaves the level, and then the second electron with $\sigma_2$ passes through it.
Finally, the third electron with $\sigma_3$ occupies it.
This process describes that the spin in the QD level fluctuates from $\sigma_1$ to $\sigma_2$ to $\sigma_3$ (from $\uparrow$ to $\uparrow$ to $\downarrow$, in this specific case shown in Fig.~\ref{001}b) and that the spin of the first electron affects that of the third electron.
The correlation among $\sigma_1$ and $\sigma_3$ via $\sigma_2$, or three-spin exchange process, gives the most naive picture of the three-body correlations, which constitute the leading term of the nonlinear response of the system as characterized by $\chi_{\sigma_1\sigma_2\sigma_3}$ ($\chi_{\uparrow\uparrow\downarrow}$ in this case).


In this work, using a carbon nanotube QD in the $SU(2)$ Kondo regime, we experimentally prove that the three-body correlations $\chi_{\sigma_1 \sigma_2 \sigma_3}$ indeed contribute to the nonlinear conductance. Thanks to the quality of our sample, where the Kondo effect in the unitary limit is achieved, we quantitatively measure the three-body correlations, in perfect agreement with recent results of the Fermi liquid theory~\cite{MoraPRB2015,FilipponePRB2018,OguriPRB2018,OguriPRL2018,TerataniPRB2020,TerataniPRL2020}, and verify their role in the non-equilibrium regime.
In particular, we demonstrate their importance when TRS is broken, and solve a long-standing puzzle of the Kondo system under the magnetic field~\cite{FilipponePRB2018}.
The demonstrated method to relate three-body correlations and  non-equilibrium transport opens up a way for further investigation of the dynamics of quantum many-body systems.

\subsection*{{\large {Results}}}

\begin{figure}
\center \includegraphics[width=\linewidth]{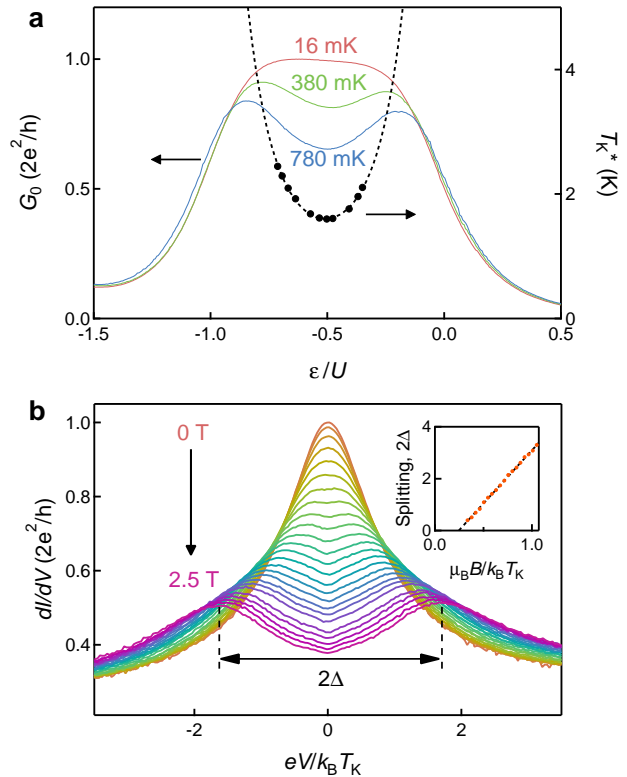}
\caption{{\bf {Energy-level and magnetic-field dependence of the Kondo state.}} {\bf a} $\varepsilon/U$ dependence of $G_{0}$ for three temperatures ($16$, $380$, and $780\ \rm{mK}$). The experimental $T_{\rm K}^{\ast}$ shown on the right axis is fitted with the formula shown in Ref.~\cite{FerrierNatPhys2016}. {\bf b} Differential conductance at the PHS point as a function of $eV/k_{\rm B}T_{\rm K}$ for different magnetic fields from $0$ to $2.5\ \rm{T}$ with $0.1\ \rm{T}$ steps. Here, the lowest field $0.1\ {\rm T}$ is defined as $0\ {\rm T}$ (see text). (inset) $2\Delta$, which is the width of Zeeman splitting, as a function of $\mu_{\rm B}B/k_{\rm B}T_{\rm K}$.}
\label{002}
\end{figure}

{\bf \noindent {Nonlinear conductance in Kondo regime.}}
We first outline the basic idea of the present study.
We examine the differential conductance ($dI/dV$) of the Kondo-correlated QD with the following expansion in terms of the bias voltage ($V$)~\cite{PustilnikJPCM2004, OguriJPSJ2005, RinconPRB2009, RinconPRB2010err, RouraBasPRB2010, SelaPRB2009, MoraPRB2009}:
\begin{eqnarray}
\dfrac{dI}{dV}=G_0-{\alpha_{\rm V}}\left(\dfrac{eV}{k_{\rm B}T_{\rm K}}\right)^2 +\cdots,\label{eqn1}
\end{eqnarray}
at $T\ll T_{\rm K}$.
Here $G_0$, $e$, and $k_{\rm B}$ are the zero-bias conductance, elementary charge, and the Boltzmann constant, respectively.
The coefficient $\alpha_{\rm V}$ of the order $V^2$ consists of two- and three-body terms, $W_2$ and $W_3$, respectively.
As discussed later, $W_2$ is defined using $\delta_\sigma$ and $\chi_{\sigma_1 \sigma_2}$ (and hence $R$) [see Equation (\ref{eqn2})], which was previously addressed and established experimentally~\cite{GrobisPRL2008, ScottPRB2009, YamauchiPRL2011, KretininPRB2011,FerrierNatPhys2016}.
In this work, we newly focus on $W_3$, which is the function of $\delta_\sigma$, $\chi_{\uparrow\uparrow}$, $\chi_{\uparrow \uparrow \uparrow}$, and $\chi_{\uparrow \uparrow \downarrow}$ for specific cases [see Equation (\ref{def_W3}), Methods, and Supplementary Note 1]~\cite{MoraPRB2015,FilipponePRB2018,OguriPRB2018,OguriPRL2018,TerataniPRB2020,TerataniPRL2020}.
In the analysis, we rely on the fact that $G_0 = e^2/h\times(\sin^2 \delta_\uparrow+\sin^2 \delta_\downarrow)$ in the Kondo effect in the unitary limit.
This treatment is valid in this experimental work as we have achieved the unitary limit in our QD (see Fig.~\ref{002}a) ~\cite{FerrierNatPhys2016, FerrierPRL2017}.

\newsavebox{\ieqnhl} \sbox{\ieqnhl}{\ref{eqn1}}
Importantly, when the on-site Coulomb interaction ($U$) in the QD  is zero, a complete analytical form for Equation ({\usebox\ieqnhl}) is obtained by replacing $k_{\rm B}T_{\rm K}$ with $2\gamma_0$ (see Methods and Supplementary Note 1).
Here, $2\gamma_0$ virtually corresponds to the half width of a resonance peak.
Note that the three-body correlations for the same spins $(\chi_{\uparrow\uparrow\uparrow}$ and $\chi_{\downarrow\downarrow\downarrow})$ are finite even in this $U=0$ case due to the Pauli exclusion principle.
On the other hand, only for the $U\neq 0$ case, the Kondo effect makes other components such as $\chi_{\uparrow\uparrow\downarrow}$ finite.
Thus, the experimental observation of the deviation from the $U=0$ case{, which we refer as the free particle (FP) model later,} directly tells us the relevance of those correlations in the nonlinear Kondo regime.
In addition, recently, the expression of the coefficient $\alpha_{\rm V}$ in the interacting case was theoretically obtained by expanding the Fermi liquid theory to the non-equilibrium regime~\cite{MoraPRB2015,FilipponePRB2018,OguriPRB2018,OguriPRL2018,TerataniPRB2020,TerataniPRL2020}.
Our present work, relying on a precise experiment, aims to perform an accurate test of this theory.

To this end, we experimentally examine $\alpha_{\rm V}$ when TRS or the particle-hole symmetry (PHS) is broken.
In the presence of the magnetic filed ($B$), a single electron energy with spin $\sigma$ in the QD is expressed as $\varepsilon_\sigma=\varepsilon-g\sigma\mu_{\rm B}B/2$, where $\mu_{\rm B}$ and $g\approx2$ are the Bohr magneton and the $g-$factor in the nanotube, respectively.
When both TRS and PHS hold, the energy level is located at $\varepsilon_\sigma=-U/2$ (center of Fig.~\ref{001}c), the situation that many experimental works have addressed~\cite{GrobisPRL2008, ScottPRB2009, YamauchiPRL2011, KretininPRB2011,FerrierNatPhys2016}.
In this symmetrical point, the two-body correlations are finite, while the three-body correlations are quenched due to the symmetry.
We systematically tune $\varepsilon_\sigma$ to investigate the behavior of $\alpha_{\rm V}$ by varying either $B$ or the gate voltage to vary $\varepsilon$.
As is known in the Kondo physics, the magnetic field splits the degenerate level by a quantity $g\mu_{\rm B}B$, which breaks TRS (see the left panel of Fig.~\ref{001}c).
The gate voltage, on the other hand, moves $\varepsilon$ from $-U/2$ and breaks PHS (see the right panel of Fig.~\ref{001}c)~\cite{MoraPRB2015,FilipponePRB2018,OguriPRB2018,OguriPRL2018}.

\newsavebox{\irefhl} \sbox{\irefhl}{\cite{FerrierNatPhys2016,FerrierPRL2017,HataArxiv2018} }
\vspace{\baselineskip}
{\bf \noindent {Sample and measurement setup.}}
We perform conductance measurement for a carbon nanotube QD connected to two Pd/Al electrodes in the dilution fridge, using a standard low-frequency lock-in technique.
We focus on the conventional $SU(2)$ Kondo state previously reported in Refs.~{\usebox\irefhl}.
The details are presented in those references and in Supplementary Note 4.
The shape of the Coulomb diamond and the shot noise measurement experimentally determines whether the symmetry of the Kondo state is $SU(2)$ or $SU(4)$~{\usebox\irefhl}.
The base temperature of the dilution fridge is $16\ {\rm mK}$.
A small magnetic field $0.1\ {\rm T}$, which is sufficiently small compared to the Kondo temperature ($T_{\rm K}=1.6\ {\rm K}$ at the PHS point), is always applied to suppress the superconductivity of the electrodes.
In this paper, for simplicity, we refer to this lowest field $0.1\ {\rm T}$ as zero field ($0\ {\rm T}$), except when we mark experimental points at $B=0.1\ {\rm T}$ in the figures.

\vspace{\baselineskip}
{\bf \noindent {Basic characteristics of the Kondo state.}}
Figure \ref{002}a shows the zero-bias conductance $G_0$ as a function of $\varepsilon/U$ at zero field for three different temperatures.
In this paper, we define $T_{\rm K}$ as the Kondo temperature at the TRS and PHS point ($B=0\ {\rm T}$ and $\varepsilon/U=-0.5$).
We also define $T_{\rm K}^{\ast}(\varepsilon, B)$ as the Kondo temperature for general cases.
By definition, $T_{\rm K}^{\ast}(\varepsilon/U=-0.5, B=0\ {\rm T})\equiv T_{\rm K}=1.6\ {\rm K}$. 
The black points in Fig.~\ref{002}a are the Kondo temperatures $T_{\rm K}^{\ast}(\varepsilon, B=0\ {\rm T})$, which we evaluate by analyzing the temperature dependence of $G_0$~\cite{FerrierNatPhys2016}.
We obtain $U=6.4 \pm 0.8\ \rm{meV}$ and $\Gamma = 1.9 \pm 0.2\ \rm{meV}$ by fitting the gate voltage dependence of $T_{\rm K}^{\ast}(\varepsilon, B=0\ {\rm T})$ (dashed curve), where $\Gamma$ is the width of energy levels due to coupling between the electron in the QD and the lead electrons~\cite{FerrierNatPhys2016}. 
Together with $U/\Gamma=3.4\pm0.6$, $R=1.95\pm0.05$, which we independently determined from the shot noise experiment~\cite{FerrierNatPhys2016}, indicates that the Kondo state is very close to the limit of the strong correlation~\cite{OguriJPSJ2005}.
As the two-body correlation $W_2$ is defined~\cite{OguriPRB2018,OguriPRL2018,TerataniPRB2020,TerataniPRL2020} by
\begin{eqnarray}
W_2=-\left[1+5\left(R-1\right)^2\right]\sum_{\sigma}\dfrac{\cos{2\delta_\sigma}}{2}.
\label{eqn2}
\end{eqnarray}
$W_2=5.5$ at $R=1.95$ and the symmetric point ($\delta_\sigma=\pi/2$).


Figure~\ref{002}b shows the differential conductance obtained at $16\ \rm{mK}$ and the PHS point ($\varepsilon/U=-0.5$) as a function of $eV/k_{\rm B}T_{\rm K}$ for different $B$ from $0$ to $2.5\ \rm{T}$.
When $B$ increases, the Kondo resonance gets split, and its amplitude is reduced. Inset in Fig.~\ref{002}b shows the Zeeman splitting $2\Delta$ as a function of $\mu_{\rm B}B/k_{\rm B}T_{\rm K}$. 
The linear fitting shows that the $g$-factor is $2.0\pm0.05$ just as expected for carbon nanotube and the splitting seems to start at $\mu_{\rm B}B/k_{\rm B}T_{\rm K}=0.23\pm0.02$ ($B=0.6\ \rm{T}$).
Although this splitting has served as a hallmark of the Kondo effect~\cite{GoldhaberGordonNature1998,Cronenwett540,SCHMID1998182,vanderWielScience2000}, the microscopic mechanism has been theoretically revealed only recently~\cite{FilipponePRB2018, OguriPRB2018, OguriPRL2018,TerataniPRB2020,TerataniPRL2020}.

\begin{figure}
\center \includegraphics[width=70mm]{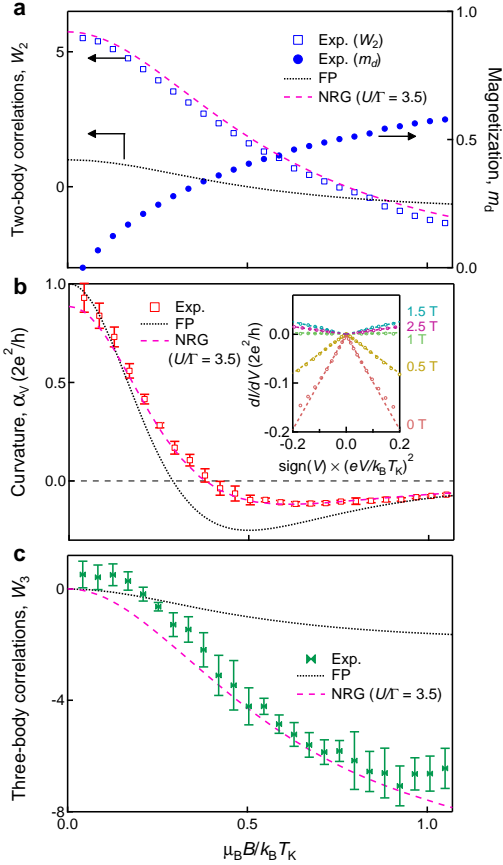}
\caption{{\bf {Two- and three-body correlations as a function of magnetic field.}} {\bf a} Two-body correlations $W_2$ and magnetization $m_d$ as a function of $\mu_{\rm B}B/k_{\rm B}T_{\rm K}$. The scaling factor $k_{\rm B}T_{\rm K}$ is $138\ {\rm \mu eV}$. $\mu_{\rm B}B/k_{\rm B}T_{\rm K}=1.0$ corresponds to $B=2.4\ {\rm T}$. The marks indicate the results deduced from the experiment. The dotted and dashed curves for $W_2$ are given by the free particle (FP) model and the NRG calculations ($U/\Gamma=3.5$), respectively. {\bf b} Curvature $\alpha_{\rm V}$ as a function of $\mu_{\rm B}B/k_{\rm B}T_{\rm K}$. The dotted and dashed curves are derived by the free particle (FP) model and the NRG calculations ($U/\Gamma=3.5$), respectively. (Inset) $dI/dV$ as a function of ${\rm sign}(V)\times(eV/k_{\rm B}T_{\rm K})^2$ at different magnetic fields. The curves are offset for clarity.  We obtained $\alpha_{\rm V}$ with linear fitting. Error bars correspond to the uncertainty of the linear fit performed on slightly different ranges. {\bf c} Three-body correlations $W_3$ as a function of $\mu_{\rm B}B/k_{\rm B}T_{\rm K}$. In the analysis, we have compared the experimental results with the FP model (dotted curves) and the NRG results for the Anderson model with $U/\Gamma=3.5$ (dashed curves)~\cite{OguriPRB2018, OguriPRL2018,TerataniPRB2020,TerataniPRL2020}. The error bars are determined using error propagation.}
\label{003}
\end{figure}

\vspace{\baselineskip}
{\bf \noindent {Two- and three-body correlations in TRS broken case.}}
We analyze the magnetic field dependence of the Kondo peak at the PHS point with Equation~(\ref{eqn1}) (the analysis procedure is detailed in the Supplementary Note 3).
At a finite magnetic field, the phase shift relates to the induced magnetization of the electron inside the QD ($m_d$) such that 
\begin{equation}
m_d=\frac{\delta_\uparrow-\delta_\downarrow}{\pi}.
\end{equation}
Then, $G_0$ and $W_2$ are expressed as:
\begin{equation}
G_0=\dfrac{2e^2}{h}\cos^2{\left(\frac{\pi m_d}{2}\right)}
\end{equation}
and
\begin{equation}
W_2=\left[1+5\left(R-1\right)^2\right]\cos{\left(\pi m_d\right)},
\end{equation}
respectively~\cite{OguriPRB2018}.
Note that to derive them we use the Friedel sum rule, $1=\left(\delta_\uparrow+\delta_\downarrow\right)/\pi,$  which holds even under the magnetic field.
The gradual increase of $m_d$ deduced from the experimental $G_0$ is shown on the right axis in Fig.~\ref{003}a.
The numerical renormalization group (NRG) calculations tell that the Wilson ratio is almost insensitive to the magnetic field up to $\mu_{\rm B}B/k_{\rm B}T_{\rm K} \sim 1$ (see Methods and Supplementary Note 2).
Thus, as shown in Fig.~\ref{003}a, the magnetic-field dependence of $W_2$ is directly deduced using the experimentally obtained $m_d$ and zero-field Wilson ratio $R=1.95$~\cite{FerrierNatPhys2016}. 
To the best of our knowledge, such a quantitative measurement of the two-body correlation of the Kondo state under magnetic fields has never been reported.

Let us now analyze the non-equilibrium case of Fig.~\ref{002}b in detail based on Equation (\ref{eqn1}).
The inset of Fig.~\ref{003}b shows $dI/dV$ as a function of ${\rm sign}(V)\times(eV/k_{\rm B}T_{\rm K})^2$ at $B=0,\ 0.5,\ 1,\ 1.5$, and $2.5\ \rm{T}$, respectively.
The ``V-shaped'' curves indicate that the conductance shows a parabolic behavior around zero bias, and importantly, the sign of their curvature changes around $1\ {\rm T}$ when $B$ increases.
This sign reversal can also define the splitting of the Kondo peak, on which we focus here, instead of the above $2\Delta$~\cite{Hewson2005}.
We obtain the curvature, that is, the coefficient $\alpha_{\rm V}$ from the fitting with Equation~(\ref{eqn1}) (see Methods and Supplementary Note 2).
Figure \ref{003}b represents $\alpha_{\rm V}$ as a function of the normalized magnetic field $\mu_{\rm B}B/k_{\rm B}T_{\rm K}$.
${\alpha_{\rm V}}$ crosses zero at $\mu_{\rm B}B/k_{\rm B}T_{\rm K}=0.38$ ($B=0.9\ \rm{T}$).

In order to elucidate the physical meaning of the magnetic field dependence of ${\alpha_{\rm V}}$, we consider the FP model with a single resonant level (see Methods and Supplementary Note 1).
In this case, analytic solutions of ${\alpha_{\rm V}}$ are given only by using a single parameter, coupling constant ($\gamma_0$).
$2\gamma_0$ corresponds to the half width of $dI/dV$ at zero field.
When $\gamma_0$ is given, the differential conductance in the magnetic field can be straightforwardly calculated for the FP case.
$\alpha_{\rm V}(B)$ thus obtained is shown in the dotted curve in Fig.~\ref{003}b.
Clearly, this dotted curve fails to explain the experimental observation, indicating the relevance of the Kondo correlation.

Now, using NRG calculations, we obtain $\alpha_{\rm V}(B)$ for the Kondo-correlated QD with $U/\Gamma =3.5$~\cite{OguriPRB2018}.
As shown in Fig.~\ref{003}b, the theoretical curve (red dashed curve) satisfactorily reproduces the experiment.

Then what can we learn from this agreement between the experiment and the theory?
As recently discussed~\cite{FilipponePRB2018,OguriPRB2018,OguriPRL2018,TerataniPRB2020,TerataniPRL2020}, $\alpha_{\rm V}$ is related to the three-body correlations $W_3$ as well as $W_2$: 
\begin{align}
\alpha_{\rm V}(B)&=\dfrac{2e^2}{h}\dfrac{\pi^2}{64}({W_2}+{W_3})\times\left(\dfrac{T_{\rm K}}{T_{\rm K}^{\ast}}\right)^2.\label{eqn3}
\end{align}
Here, $W_3$ is defined as follows (see Methods and Supplementary Note 1), 
\begin{equation}
W_3=\frac{1}{2\chi_{\uparrow\uparrow}^2}\sum_{\sigma}\dfrac{\sin{2\delta_\sigma}}{2\pi}\left(\chi_{\sigma\sigma\sigma}+3\chi_{\sigma-\sigma-\sigma}\right). 
\label{def_W3}
\end{equation}
As shown in Fig.~\ref{003}a, we have already obtained the magnetic field dependence of $W_2$.
In addition, $T_{\rm K}^{\ast}(\varepsilon=0,B)$ as a function of $B$ is obtained from the Lorentzian fitting of $dI/dV$ (see Supplementary Note 2).
Thus, we can finally derive the magnetic-field dependence of $W_3$ using Equation~(\ref{eqn3}) as presented in Fig.~\ref{003}c.
$W_3$ is zero at zero field and becomes finite as $B$ is increased, i.e., as the TRS is broken.

We can also analytically derive $W_3$ in the FP model as shown in the dotted curve in Fig.~\ref{003}c [see Methods and Supplementary Note 1].
Importantly, as is clear in this figure, the absolute value of the experimental $W_3$ is much larger than the theoretical value of the FP model.
The behavior of $W_3$ is clearly due to the enhancement of the three-body correlations $\chi_{\sigma_1\sigma_2\sigma_3}$ by the Kondo effect.
Indeed, NRG calculations with $U/\Gamma=3.5$ shown in a red dashed curve nicely reproduce the experimental $W_3$.
This result assures that the residual interaction manifests itself in the non-equilibrium transport in the symmetry-breaking regime.
The experimental determination of the three-body correlations is our central achievement of this work.

\begin{figure}
\center \includegraphics[width=70mm]{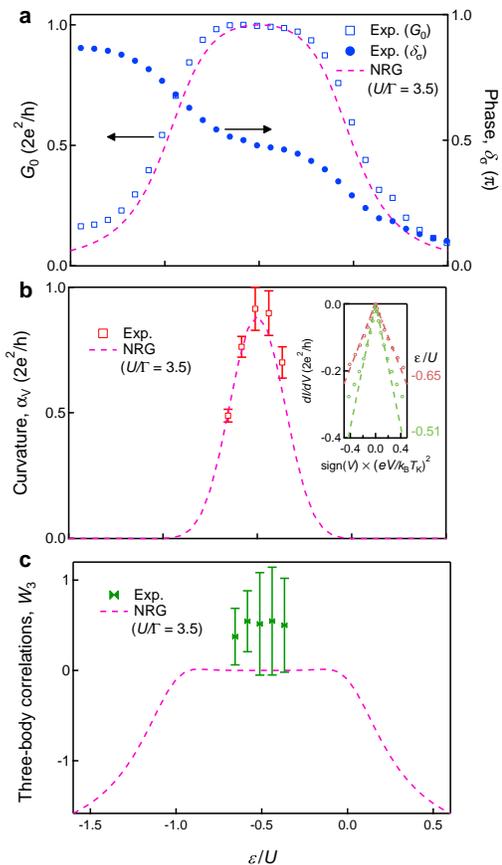}
\caption{{\bf {Two- and three-body correlations as a function of the energy level.}} {\bf a} Zero-bias conductance $G_{0}$ and phase shift $\delta_\sigma$ as a function of $\varepsilon/U$. The marks are experimental results obtained at $B=0$. The dashed curve for $G_0$ is given by the NRG calculations with $U/\Gamma=3.5$. {\bf b} Curvature $\alpha_{\rm V}$ as a function of $\varepsilon/U$.  The dashed curve is the NRG results with $U/\Gamma=3.5$. (Inset) $dI/dV$ as a function of ${\rm sign}(V)\times(eV/k_{\rm B}T_{\rm K})^2$ at $\varepsilon/U=-0.51$ and $-0.65$. Error bars correspond to the uncertainty of the linear fit performed on slightly different ranges. {\bf c} Three-body correlations $W_3$ as a function of $\varepsilon/U$. We compare the experimental results with the NRG results with $U/\Gamma=3.5$ (dashed curve)~\cite{OguriPRB2018, OguriPRL2018,TerataniPRB2020,TerataniPRL2020}. The error bars are determined using error propagation.}
\label{004}
\end{figure}

\vspace{\baselineskip}
{\bf {Two- and three-body correlations in PHS broken case.}}
So far, we have focused on the PHS point ($\varepsilon/U=-0.5$, see Fig.~\ref{001}c).
Next, we address the PHS breaking regime by tuning the gate voltage ($\varepsilon$) at $B=0$.
The square mark in Fig.~\ref{004}a shows $\varepsilon/U$ dependence of $G_0$, which decreases as $\varepsilon$ deviates from the PHS point.
The dashed curve given by the NRG calculations again agrees with the experimental results well.
We also show the phase shift, $\delta_\sigma (=\delta_\uparrow = \delta_\downarrow)$, evaluated from Equation (\ref{eqn1})  (circles in Fig.~\ref{004}a).
As expected, the phase is locked around $\pi/2$ and develops a plateau~\cite{TakadaPRL2014,TakadaPRB2016}.
The NRG calculations indicate that the Wilson ratio does not decrease in this region (see Supplementary Note 2).
Relying on this fact, we evaluate the $\varepsilon/U$-dependence of the two-body correlation $W_2$ with Equation (\ref{eqn2}) and the experimental value of the Wilson ratio $R=1.95$ at $B=0$ (For the obtained $W_2$, see Supplementary Figure 5).

The inset of Fig.~\ref{004}b shows $dI/dV$ as a function of ${\rm sign}(V)\times(eV/k_{\rm B}T_{\rm K})^2$ at $\varepsilon/U=-0.51$ and $-0.65$.
The slope at the former (around the PHS point) is larger than that at the latter (PHS broken). 
Figure \ref{004}b is the deduced curvature $\alpha_{\rm V}$ as a function of $\varepsilon/U$.
Here, we investigate the region within $|\varepsilon/U+0.5|<0.15$, where we experimentally obtain the Kondo temperature, $T_{\rm K}^{\ast}(\varepsilon,B=0\ {\rm T})$ (see Fig.~\ref{002}a).
In Fig.~\ref{004}c, we plot $W_3$ evaluated with Equation (\ref{eqn3}), showing that $W_3$ remains almost zero in the entire region of $|\varepsilon/U+0.5|<0.15$, which is in sharp contrast with the above result in the magnetic field.
This indicates that three-body correlations are more sensitive to magnetic field modulation than the gate voltage modulation (energy level modulation). This reflects that the spin degree of freedom of the Kondo state is fluctuating while the charge part is frozen.
The NRG calculations with $U/\Gamma=3.5$ shown in a dashed curve agree with the experimental $\alpha_{\rm V}$ (Fig.~\ref{004}b) and $W_3$ (Fig.~\ref{004}c).
This agreement again guarantees that the three-body correlations can be estimated experimentally.


\subsection*{{\large {Discussion}}}
In conclusion, we report the experimental determination of  many-body correlations in the Kondo QD.
We demonstrate that the curvatures of the differential conductance in the TRS- and PHS- broken regimes are well reproduced by the NRG calculations with strong interactions, enabling us to successfully evaluate the two-body and three-body correlations.
The demonstrated quantitative analysis confirms the validity of the recent development of the Fermi liquid theory in the non-equilibrium regime, which should inspire future work, for example, to cover the temperature region around and above $T_{\rm K}$.
In particular, their sensitivity to TRS and PHS should make the three-body correlations an efficient tool to explore correlated quantum liquids such as topological spin liquids.


\section*{{Methods}}
{\bf \noindent Fabrication process}
We show the outline of the fabrication process~\cite{DelagPhD}.
\begin{enumerate}
  \item Sputtering Fe catalyst with $1\ \rm{nm}$ thickness on intrinsic Si substrate.
  \item Placing the substrate in a quartz tube and heat it in a oven with low pressure ($\sim 1.1\times 10^{-4}\ \rm{mbar}$) and stabilized temperature between $800$ and $1000\ \rm{^\circ C}$.
  \item Injecting $\rm{C_{2}H_{2}}$ into the quartz tube for 9 seconds and pumping it out. After this process, we can get carbon nanotubes.
  \item Connecting the carbon nanotubes with metallic leads by the electron beam lithography. We used Pd/Al bilayer for electrodes in the experiment [Supplementary Figure 6 (a)].
\end{enumerate}

\vspace{\baselineskip}
{\bf \noindent Analysis in the $U=0$ case}
The complete analytical form for Equation~(\ref{eqn1}) is obtained in the $U=0$ case (see Supplementary Note 1), which is used in the FP model in this work.
The experimental Kondo peak at zero field is treated as if it were a resonance peak with $U=0$. We obtain $\gamma_0=65\ {\rm \mu eV}$ by fitting the peak with the following formula:
\begin{align}
\dfrac{dI}{dV}=\dfrac{2e^2}{h}\dfrac{\gamma_0^2}{(eV/2)^2+\gamma_0^2}.
\end{align}
As $k_{\rm B}T_{\rm K}=138\ {\rm \mu eV}$, we see $k_{\rm B}T_{\rm K}/\gamma_0\sim2$, which is explained in Supplementary Note 1.

We present the magnetic field dependence of $W_2$, $\alpha_{\rm V}$, and $W_3$ in Figs.~\ref{003}a, 3b, and 3c, respectively, by using:
\begin{align}
W_{2}&=\dfrac{\gamma_0^2-(\mu_{\rm B}B)^2}{(\mu_{\rm B}B)^2+\gamma_0^2},\\
\alpha_{\rm V}&=\dfrac{2e^2}{h}\dfrac{1-3(\mu_{\rm B}B/\gamma_0)^2}{\left\{1+(\mu_{\rm B}B/\gamma_0)^2\right\}^3},\\
W_{3}&=\dfrac{-2(\mu_{\rm B}B)^2}{(\mu_{\rm B}B)^2+\gamma_0^2}.
\label{eqn013}
\end{align}
We treat $\mu_{\rm B}B/\gamma_0= (k_{\rm B}T_{\rm K}/\gamma_0) \times (\mu_{\rm B}B/k_{\rm B}T_{\rm K})$. In Figs.~\ref{003}a, 3b, and 3c, $\mu_{\rm B}B/k_{\rm B}T_{\rm K}$ is taken as the horizontal axis.

\vspace{\baselineskip}
{\bf \noindent Properties of $\chi_{\sigma_1 \sigma_2 \sigma_3}$}
Three-body correlations have permutation symmetry for the spin indexes: 
\begin{align}
  \begin{split}
  &\chi_{\sigma_1 \sigma_2 \sigma_3}=\chi_{\sigma_2 \sigma_3 \sigma_1}=\chi_{\sigma_3 \sigma_1 \sigma_2}\\
  &=\chi_{\sigma_1 \sigma_3 \sigma_2}
  =\chi_{\sigma_2 \sigma_1 \sigma_3}=\chi_{\sigma_3 \sigma_2 \sigma_1}. 
  \end{split}
\end{align}
At zero magnetic field (the TRS point), $\chi_{\uparrow \uparrow \uparrow}=\chi_{\downarrow \downarrow \downarrow}$ and $\chi_{\uparrow \downarrow \downarrow}=\chi_{\uparrow \uparrow \downarrow}$ hold. At the PHS point, $\chi_{\uparrow \uparrow \uparrow}=-\chi_{\downarrow \downarrow \downarrow}$ and $\chi_{\uparrow \downarrow \downarrow}=-\chi_{\uparrow \uparrow \downarrow}$ hold. These properties are used in the discussion of $W_3$ defined by Equation (\ref{def_W3}).

\vspace{\baselineskip}
{\bf \noindent Analysis of differential conductance}
In the analysis shown in Fig.~\ref{002}b, we do not take the data obtained at very low bias region (typically $|eV/k_{B}T_{K}|^2<0.01$), where a slight deviation from the parabolicity is observed. This effect is most probably due to some other lower energy physics such as two-stage Kondo effect (see Supplementary Note 2 and Ref.~\cite{vanderWielPRL2002}).

\vspace{\baselineskip}
{\bf \noindent Numerical calculations}
NRG results shown in Figs.\ 3 and 4 have been calculated using 
the Anderson impurity model in the form  $\mathcal{H} = 
\mathcal{H}_{d} + \mathcal{H}_{c} + \mathcal{H}_{T}$,
\begin{align*}
\mathcal{H}_{d}\,&= \sum_{\sigma=\uparrow,\downarrow}^{}\,
\varepsilon_{\sigma}^{} \,d_{\sigma}^{\dagger}d_{\sigma}^{}
\,+\,U\,d_{\uparrow}^{\dagger}d_{\uparrow}^{}\,
d_{\downarrow}^{\dagger}d_{\downarrow}^{}\,, 
\\
\mathcal{H}_{c}\,&= 
\sum_{\sigma=\uparrow,\downarrow}^{} 
\sum_{\lambda=L,R}
\int_{-D}^D  \! d\xi\,  \xi\, 
c^{\dagger}_{\xi \lambda \sigma} c_{\xi \lambda \sigma}^{}\,, 
\\
\mathcal{H}_{T}\,&=  \sum_{\sigma=\uparrow,\downarrow}^{}  
\sum_{\lambda=L,R} v_{\lambda}^{}
\int_{-D}^D d\xi \,\sqrt{\rho_c^{}}  
 \left( c^{\dagger}_{\xi \lambda \sigma}  d_{\sigma}^{}
+ \mathrm{H.c.} \right) .
\end{align*}
Here,  $\varepsilon_{\sigma}^{}=\varepsilon -\sigma \mu_B B$ is the energy of  a single electron with spin $\sigma$ and $U$ is the Coulomb interaction between electrons  in the QD. 
The $g$-factor is taken as 2.
$\mathcal{H}_{c}$ describes the conduction bands of the leads on the left ($L$) and right ($R$) with  the constant density of states  $\rho_{c}\equiv 1/(2D)$. 
When $U=0$, the resonance width of the local level due to the tunnel couplings in $\mathcal{H}_{T}$ is given by $\Gamma/2 = \pi \rho_{c} \left( v_{L}^{2} + v_{R}^{2} \right)$. 
We assume that the couplings to be symmetric $v_L=v_R$. 

The NRG calculations have been carried out choosing the discretization parameter to be  $\Lambda=2.0$ and keeping $N_\mathrm{trunc}=3600$ low-lying energy states at each step of the iterative procedure~\cite{Bulla2008}.
Our NRG code uses the global $\mathrm{U}\left(1\right)\otimes\mathrm{SU}\left(2\right)$ symmetries and the method for deducing the correlation functions described in Ref.~\cite{TerataniPRL2020}.

\vspace{\baselineskip}
{\bf \noindent {Data availability}}
The data that support the findings of this study are available from the corresponding author upon reasonable request.



\vspace{\baselineskip}
{\bf \noindent {Acknowledgment}}
We thank Rapha\"elle Delagrange for helpful discussions and for sample fabrication.
K.K. acknowledges the stimulating discussion in the meeting of the
Cooperative Research Project of the RIEC, Tohoku University.
This work was partially supported by JSPS KAKENHI Grant (No.~JP19K14630, No.~JP19H05826, No.~JP19H00656, No.~JP18K03495, JP18H01815, and No.~JP18J10205), JST CREST Grant (No.~JPMJCR1876), and the French program ANR JETS (ANR-16-CE30-0029-01).

This is a preprint of an article published in Nature Communications.
The final authenticated version is available online at: \url{https://doi.org/10.1038/s41467-021-23467-4}.

\vspace{\baselineskip}
{\bf \noindent {Author contributions}}
M.F. and R.D. fabricated the sample. T.H., M.F., T.A., and S.L. performed the experiments. T.H. analysed the data. Y.T., R.S., and A.O. performed the theoretical simulations. T.H. and K.K. wrote the paper with the input of all authors. K.K. and A.O. supervised the research.

\vspace{\baselineskip}
{\bf \noindent {Competing interests}}
The authors declare no competing interests.


\begin{thebibliography}{1}
\expandafter\ifx\csname url\endcsname\relax
  \def\url#1{\texttt{#1}}\fi
\expandafter\ifx\csname urlprefix\endcsname\relax\def\urlprefix{URL }\fi
\providecommand{\bibinfo}[2]{#2}
\providecommand{\eprint}[2][]{\url{#2}}

\bibitem{OguriPRL2018}
\bibinfo{author}{Oguri, A.} \& \bibinfo{author}{Hewson, A.~C.}
\newblock \bibinfo{title}{Higher-order {Fermi}-liquid corrections for an
  {Anderson} impurity away from half filling}.
\newblock \emph{\bibinfo{journal}{Physical Review Letters}}
  \textbf{\bibinfo{volume}{120}}, \bibinfo{pages}{126802}
  (\bibinfo{year}{2018}).

\bibitem{NozieresJLTP1974}
\bibinfo{author}{Nozieres, P.}
\newblock \bibinfo{title}{A {"Fermi-liquid"} description of the {Kondo} problem
  at low temperatures}.
\newblock \emph{\bibinfo{journal}{Journal of Low Temperature Physics}}
  \textbf{\bibinfo{volume}{17}}, \bibinfo{pages}{31--42}
  (\bibinfo{year}{1974}).

\bibitem{YosidaPTPS1970}
\bibinfo{author}{Yosida, K.} \& \bibinfo{author}{Yamada, K.}
\newblock \bibinfo{title}{Perturbation expansion for the anderson hamiltonian}.
\newblock \emph{\bibinfo{journal}{Progress of Theoretical Physics Supplement}}
  \textbf{\bibinfo{volume}{46}}, \bibinfo{pages}{244--255}
  (\bibinfo{year}{1970}).

\bibitem{OguriJPSJ2005}
\bibinfo{author}{Oguri, A.}
\newblock \bibinfo{title}{Fermi liquid theory for the nonequilibrium {Kondo}
  effect at low bias voltages}.
\newblock \emph{\bibinfo{journal}{Journal of the Physical Society of Japan}}
  \textbf{\bibinfo{volume}{74}}, \bibinfo{pages}{110--117}
  (\bibinfo{year}{2005}).

\bibitem{OguriPRB2018_2}
\bibinfo{author}{Oguri, A.} \& \bibinfo{author}{Hewson, A.~C.}
\newblock \bibinfo{title}{Higher-order {Fermi}-liquid corrections for an
  {Anderson} impurity away from half filling: {Nonequilibrium} transport}.
\newblock \emph{\bibinfo{journal}{Physical Review B}}
  \textbf{\bibinfo{volume}{97}}, \bibinfo{pages}{035435}
  (\bibinfo{year}{2018}).

\bibitem{vanderWielPRL2002}
\bibinfo{author}{van~der Wiel, W.~G.} \emph{et~al.}
\newblock \bibinfo{title}{Two-stage {Kondo} effect in a quantum dot at a high
  magnetic field}.
\newblock \emph{\bibinfo{journal}{Physical Review Letters}}
  \textbf{\bibinfo{volume}{88}}, \bibinfo{pages}{126803}
  (\bibinfo{year}{2002}).

\bibitem{FerrierNatPhys2016}
\bibinfo{author}{Ferrier, M.} \emph{et~al.}
\newblock \bibinfo{title}{Universality of non-equilibrium fluctuations in
  strongly correlated quantum liquids}.
\newblock \emph{\bibinfo{journal}{Nature Physics}}
  \textbf{\bibinfo{volume}{12}}, \bibinfo{pages}{230--235}
  (\bibinfo{year}{2016}).

\bibitem{FerrierPRL2017}
\bibinfo{author}{Ferrier, M.} \emph{et~al.}
\newblock \bibinfo{title}{Quantum fluctuations along symmetry crossover in a
  {Kondo}-correlated quantum dot}.
\newblock \emph{\bibinfo{journal}{Physical Review Letters}}
  \textbf{\bibinfo{volume}{118}}, \bibinfo{pages}{196803}
  (\bibinfo{year}{2017}).

\bibitem{Hata2018}
\bibinfo{author}{Hata, T.} \emph{et~al.}
\newblock \bibinfo{title}{Enhanced shot noise of multiple andreev reflections
  in a carbon nanotube quantum dot in {SU}(2) and {SU}(4) kondo regimes}.
\newblock \emph{\bibinfo{journal}{Physical Review Letters}}
  \textbf{\bibinfo{volume}{121}} (\bibinfo{year}{2018}).

\end{thebibliography}


\begin{thebibliography}{10}
\expandafter\ifx\csname url\endcsname\relax
  \def\url#1{\texttt{#1}}\fi
\expandafter\ifx\csname urlprefix\endcsname\relax\def\urlprefix{URL }\fi
\providecommand{\bibinfo}[2]{#2}
\providecommand{\eprint}[2][]{\url{#2}}

\bibitem{Axilrod1943}
\bibinfo{author}{Axilrod, B.~M.} \& \bibinfo{author}{Teller, E.}
\newblock \bibinfo{title}{Interaction of the van der {Waals} type between three
  atoms}.
\newblock \emph{\bibinfo{journal}{The Journal of Chemical Physics}}
  \textbf{\bibinfo{volume}{11}}, \bibinfo{pages}{299--300}
  (\bibinfo{year}{1943}).

\bibitem{Otsuka2010}
\bibinfo{author}{Otsuka, T.}, \bibinfo{author}{Suzuki, T.},
  \bibinfo{author}{Holt, J.~D.}, \bibinfo{author}{Schwenk, A.} \&
  \bibinfo{author}{Akaishi, Y.}
\newblock \bibinfo{title}{Three-body forces and the limit of oxygen isotopes}.
\newblock \emph{\bibinfo{journal}{Physical Review Letters}}
  \textbf{\bibinfo{volume}{105}}, \bibinfo{pages}{032501}
  (\bibinfo{year}{2010}).

\bibitem{Efimov}
\bibinfo{author}{Efimov, V.}
\newblock \bibinfo{title}{Energy levels arising from resonant two-body forces
  in a three-body system}.
\newblock \emph{\bibinfo{journal}{Physics Letters B}}
  \textbf{\bibinfo{volume}{33}}, \bibinfo{pages}{563 -- 564}
  (\bibinfo{year}{1970}).

\bibitem{Kraemer2006}
\bibinfo{author}{Kraemer, T.} \emph{et~al.}
\newblock \bibinfo{title}{{Evidence for Efimov quantum states in an ultracold
  gas of caesium atoms}}.
\newblock \emph{\bibinfo{journal}{Nature}} \textbf{\bibinfo{volume}{440}},
  \bibinfo{pages}{315--318} (\bibinfo{year}{2006}).

\bibitem{Thouless_1965}
\bibinfo{author}{Thouless, D.~J.}
\newblock \bibinfo{title}{Exchange in solid $^{3}\mathrm{He}$ and the
  {Heisenberg Hamiltonian}}.
\newblock \emph{\bibinfo{journal}{Proceedings of the Physical Society}}
  \textbf{\bibinfo{volume}{86}}, \bibinfo{pages}{893--904}
  (\bibinfo{year}{1965}).

\bibitem{Roger1983}
\bibinfo{author}{Roger, M.}, \bibinfo{author}{Hetherington, J.~H.} \&
  \bibinfo{author}{Delrieu, J.~M.}
\newblock \bibinfo{title}{Magnetism in solid $^{3}\mathrm{He}$}.
\newblock \emph{\bibinfo{journal}{Reviews of Modern Physics}}
  \textbf{\bibinfo{volume}{55}}, \bibinfo{pages}{1--64} (\bibinfo{year}{1983}).

\bibitem{Lacroix2011}
\bibinfo{author}{Lacroix, C.}, \bibinfo{author}{Mendels, P.} \&
  \bibinfo{author}{Mila, F.}
\newblock \emph{\bibinfo{title}{Introduction to Frustrated Magnetism:
  Materials, Experiments, Theory}}.
\newblock Springer Series in Solid-State Sciences (\bibinfo{publisher}{Springer
  Berlin Heidelberg}, \bibinfo{year}{2011}).
\newblock \urlprefix\url{https://books.google.co.jp/books?id=utSV09ZuhOkC}.

\bibitem{GoldhaberGordonNature1998}
\bibinfo{author}{Goldhaber-Gordon, D.} \emph{et~al.}
\newblock \bibinfo{title}{Kondo effect in a single-electron transistor}.
\newblock \emph{\bibinfo{journal}{Nature}} \textbf{\bibinfo{volume}{391}},
  \bibinfo{pages}{156--159} (\bibinfo{year}{1998}).

\bibitem{Cronenwett540}
\bibinfo{author}{Cronenwett, S.~M.}, \bibinfo{author}{Oosterkamp, T.~H.} \&
  \bibinfo{author}{Kouwenhoven, L.~P.}
\newblock \bibinfo{title}{A tunable {Kondo} effect in quantum dots}.
\newblock \emph{\bibinfo{journal}{Science}} \textbf{\bibinfo{volume}{281}},
  \bibinfo{pages}{540--544} (\bibinfo{year}{1998}).

\bibitem{SCHMID1998182}
\bibinfo{author}{Schmid, J.}, \bibinfo{author}{Weis, J.},
  \bibinfo{author}{Eberl, K.} \& \bibinfo{author}{Klitzing, K.~v.}
\newblock \bibinfo{title}{A quantum dot in the limit of strong coupling to
  reservoirs}.
\newblock \emph{\bibinfo{journal}{Physica B: Condensed Matter}}
  \textbf{\bibinfo{volume}{256-258}}, \bibinfo{pages}{182 -- 185}
  (\bibinfo{year}{1998}).

\bibitem{vanderWielScience2000}
\bibinfo{author}{van~der Wiel, W.~G.} \emph{et~al.}
\newblock \bibinfo{title}{The {Kondo} effect in the unitary limit}.
\newblock \emph{\bibinfo{journal}{Science}} \textbf{\bibinfo{volume}{289}},
  \bibinfo{pages}{2105--2108} (\bibinfo{year}{2000}).

\bibitem{NozieresJLTP1974}
\bibinfo{author}{Nozi\`{e}res, P.}
\newblock \bibinfo{title}{A {``Fermi-liquid''} description of the {Kondo}
  problem at low temperatures}.
\newblock \emph{\bibinfo{journal}{Journal of Low Temperature Physics}}
  \textbf{\bibinfo{volume}{17}}, \bibinfo{pages}{31--42}
  (\bibinfo{year}{1974}).

\bibitem{YosidaPTPS1970}
\bibinfo{author}{Yosida, K.} \& \bibinfo{author}{Yamada, K.}
\newblock \bibinfo{title}{Perturbation expansion for the {Anderson}
  {Hamiltonian}}.
\newblock \emph{\bibinfo{journal}{Progress of Theoretical Physics Supplement}}
  \textbf{\bibinfo{volume}{46}}, \bibinfo{pages}{244--255}
  (\bibinfo{year}{1970}).

\bibitem{YamadaPTP1975_1}
\bibinfo{author}{Yamada, K.}
\newblock \bibinfo{title}{Perturbation expansion for the {Anderson}
  {Hamiltonian}. {II}}.
\newblock \emph{\bibinfo{journal}{Progress of Theoretical Physics}}
  \textbf{\bibinfo{volume}{53}}, \bibinfo{pages}{970--986}
  (\bibinfo{year}{1975}).

\bibitem{YamadaPTP1975_2}
\bibinfo{author}{Yamada, K.}
\newblock \bibinfo{title}{Perturbation expansion for the {Anderson}
  {Hamiltonian. IV}}.
\newblock \emph{\bibinfo{journal}{Progress of Theoretical Physics}}
  \textbf{\bibinfo{volume}{54}}, \bibinfo{pages}{316--324}
  (\bibinfo{year}{1975}).

\bibitem{ShibaPTP1975}
\bibinfo{author}{Shiba, H.}
\newblock \bibinfo{title}{The korringa relation for the impurity nuclear
  spin-lattice relaxation in dilute {Kondo} alloys}.
\newblock \emph{\bibinfo{journal}{Progress of Theoretical Physics}}
  \textbf{\bibinfo{volume}{54}}, \bibinfo{pages}{967--981}
  (\bibinfo{year}{1975}).

\bibitem{YoshimoriPTP1976}
\bibinfo{author}{Yoshimori, A.}
\newblock \bibinfo{title}{Perturbation analysis on orbital-degenerate
  {Anderson} model}.
\newblock \emph{\bibinfo{journal}{Progress of Theoretical Physics}}
  \textbf{\bibinfo{volume}{55}}, \bibinfo{pages}{67--80}
  (\bibinfo{year}{1976}).

\bibitem{MoraPRB2015}
\bibinfo{author}{Mora, C.}, \bibinfo{author}{Moca, C.~P.}, \bibinfo{author}{von
  Delft, J.} \& \bibinfo{author}{Zarand, G.}
\newblock \bibinfo{title}{Fermi-liquid theory for the single-impurity
  {Anderson} model}.
\newblock \emph{\bibinfo{journal}{Physical Review B}}
  \textbf{\bibinfo{volume}{92}}, \bibinfo{pages}{075120}
  (\bibinfo{year}{2015}).

\bibitem{FilipponePRB2018}
\bibinfo{author}{Filippone, M.}, \bibinfo{author}{Moca, C.~P.},
  \bibinfo{author}{Weichselbaum, A.}, \bibinfo{author}{von Delft, J.} \&
  \bibinfo{author}{Mora, C.}
\newblock \bibinfo{title}{At which magnetic field, exactly, does the {Kondo}
  resonance begin to split? {A} {Fermi} liquid description of the low-energy
  properties of the {Anderson} model}.
\newblock \emph{\bibinfo{journal}{Physical Review B}}
  \textbf{\bibinfo{volume}{98}}, \bibinfo{pages}{075404}
  (\bibinfo{year}{2018}).

\bibitem{OguriPRB2018}
\bibinfo{author}{Oguri, A.} \& \bibinfo{author}{Hewson, A.~C.}
\newblock \bibinfo{title}{Higher-order {Fermi}-liquid corrections for an
  {Anderson} impurity away from half filling: Nonequilibrium transport}.
\newblock \emph{\bibinfo{journal}{Physical Review B}}
  \textbf{\bibinfo{volume}{97}}, \bibinfo{pages}{035435}
  (\bibinfo{year}{2018}).

\bibitem{OguriPRL2018}
\bibinfo{author}{Oguri, A.} \& \bibinfo{author}{Hewson, A.~C.}
\newblock \bibinfo{title}{Higher-order {Fermi}-liquid corrections for an
  {Anderson} impurity away from half filling}.
\newblock \emph{\bibinfo{journal}{Physical Review Letters}}
  \textbf{\bibinfo{volume}{120}}, \bibinfo{pages}{126802}
  (\bibinfo{year}{2018}).

\bibitem{TerataniPRB2020}
\bibinfo{author}{Teratani, Y.} \emph{et~al.}
\newblock \bibinfo{title}{Field-induced su(4) to su(2) kondo crossover in a
  half-filling nanotube dot: Spectral and finite-temperature properties}.
\newblock \emph{\bibinfo{journal}{Physical Review B}}
  \textbf{\bibinfo{volume}{102}}, \bibinfo{pages}{165106}
  (\bibinfo{year}{2020}).

\bibitem{TerataniPRL2020}
\bibinfo{author}{Teratani, Y.}, \bibinfo{author}{Sakano, R.} \&
  \bibinfo{author}{Oguri, A.}
\newblock \bibinfo{title}{Fermi liquid theory for nonlinear transport through a
  multilevel anderson impurity}.
\newblock \emph{\bibinfo{journal}{Physical Review Letters}}
  \textbf{\bibinfo{volume}{125}}, \bibinfo{pages}{216801}
  (\bibinfo{year}{2020}).

\bibitem{PustilnikJPCM2004}
\bibinfo{author}{Pustilnik, M.} \& \bibinfo{author}{Glazman, L.}
\newblock \bibinfo{title}{Kondo effect in quantum dots}.
\newblock \emph{\bibinfo{journal}{Journal of Physics: Condensed Matter}}
  \textbf{\bibinfo{volume}{16}}, \bibinfo{pages}{R513--R537}
  (\bibinfo{year}{2004}).

\bibitem{OguriJPSJ2005}
\bibinfo{author}{Oguri, A.}
\newblock \bibinfo{title}{Fermi liquid theory for the nonequilibrium {Kondo}
  effect at low bias voltages}.
\newblock \emph{\bibinfo{journal}{Journal of the Physical Society of Japan}}
  \textbf{\bibinfo{volume}{74}}, \bibinfo{pages}{110--117}
  (\bibinfo{year}{2005}).

\bibitem{RinconPRB2009}
\bibinfo{author}{Rincon, J.}, \bibinfo{author}{Aligia, A.~A.} \&
  \bibinfo{author}{Hallberg, K.}
\newblock \bibinfo{title}{Universal scaling in nonequilibrium transport through
  an {Anderson} impurity}.
\newblock \emph{\bibinfo{journal}{Physical Review B}}
  \textbf{\bibinfo{volume}{79}}, \bibinfo{pages}{121301}
  (\bibinfo{year}{2009}).

\bibitem{RinconPRB2010err}
\bibinfo{author}{Rincon, J.}, \bibinfo{author}{Aligia, A.~A.} \&
  \bibinfo{author}{Hallberg, K.}
\newblock \bibinfo{title}{Erratum: Universal scaling in nonequilibrium
  transport through an {Anderson} impurity}.
\newblock \emph{\bibinfo{journal}{Physical Review B}}
  \textbf{\bibinfo{volume}{81}}, \bibinfo{pages}{039901}
  (\bibinfo{year}{2010}).

\bibitem{RouraBasPRB2010}
\bibinfo{author}{Roura-Bas, P.}
\newblock \bibinfo{title}{Universal scaling in transport out of equilibrium
  through a single quantum dot using the noncrossing approximation}.
\newblock \emph{\bibinfo{journal}{Physical Review B}}
  \textbf{\bibinfo{volume}{81}}, \bibinfo{pages}{155327}
  (\bibinfo{year}{2010}).

\bibitem{SelaPRB2009}
\bibinfo{author}{Sela, E.} \& \bibinfo{author}{Malecki, J.}
\newblock \bibinfo{title}{Nonequilibrium conductance of asymmetric nanodevices
  in the {Kondo} regime}.
\newblock \emph{\bibinfo{journal}{Physical Review B}}
  \textbf{\bibinfo{volume}{80}}, \bibinfo{pages}{233103}
  (\bibinfo{year}{2009}).

\bibitem{MoraPRB2009}
\bibinfo{author}{Mora, C.}, \bibinfo{author}{Vitushinsky, P.},
  \bibinfo{author}{Leyronas, X.}, \bibinfo{author}{Clerk, A.~A.} \&
  \bibinfo{author}{Le~Hur, K.}
\newblock \bibinfo{title}{Theory of nonequilibrium transport in the {SU(N)}
  {Kondo} regime}.
\newblock \emph{\bibinfo{journal}{Physical Review B}}
  \textbf{\bibinfo{volume}{80}}, \bibinfo{pages}{155322}
  (\bibinfo{year}{2009}).

\bibitem{GrobisPRL2008}
\bibinfo{author}{Grobis, M.}, \bibinfo{author}{Rau, I.~G.},
  \bibinfo{author}{Potok, R.~M.}, \bibinfo{author}{Shtrikman, H.} \&
  \bibinfo{author}{Goldhaber-Gordon, D.}
\newblock \bibinfo{title}{Universal scaling in nonequilibrium transport through
  a single channel {Kondo} dot}.
\newblock \emph{\bibinfo{journal}{Physical Review Letters}}
  \textbf{\bibinfo{volume}{100}}, \bibinfo{pages}{246601}
  (\bibinfo{year}{2008}).

\bibitem{ScottPRB2009}
\bibinfo{author}{Scott, G.~D.}, \bibinfo{author}{Keane, Z.~K.},
  \bibinfo{author}{Ciszek, J.~W.}, \bibinfo{author}{Tour, J.~M.} \&
  \bibinfo{author}{Natelson, D.}
\newblock \bibinfo{title}{Universal scaling of nonequilibrium transport in the
  {Kondo} regime of single molecule devices}.
\newblock \emph{\bibinfo{journal}{Physical Review B}}
  \textbf{\bibinfo{volume}{79}}, \bibinfo{pages}{165413}
  (\bibinfo{year}{2009}).

\bibitem{YamauchiPRL2011}
\bibinfo{author}{Yamauchi, Y.} \emph{et~al.}
\newblock \bibinfo{title}{Evolution of the {Kondo} effect in a quantum dot
  probed by shot noise}.
\newblock \emph{\bibinfo{journal}{Physical Review Letters}}
  \textbf{\bibinfo{volume}{106}}, \bibinfo{pages}{176601}
  (\bibinfo{year}{2011}).

\bibitem{KretininPRB2011}
\bibinfo{author}{Kretinin, A.~V.} \emph{et~al.}
\newblock \bibinfo{title}{Spin-$\frac{1}{2}$ {Kondo} effect in an {InAs}
  nanowire quantum dot: Unitary limit, conductance scaling, and {Zeeman}
  splitting}.
\newblock \emph{\bibinfo{journal}{Physical Review B}}
  \textbf{\bibinfo{volume}{84}}, \bibinfo{pages}{245316}
  (\bibinfo{year}{2011}).

\bibitem{FerrierNatPhys2016}
\bibinfo{author}{Ferrier, M.} \emph{et~al.}
\newblock \bibinfo{title}{Universality of non-equilibrium fluctuations in
  strongly correlated quantum liquids}.
\newblock \emph{\bibinfo{journal}{Nature Physics}}
  \textbf{\bibinfo{volume}{12}}, \bibinfo{pages}{230--235}
  (\bibinfo{year}{2016}).

\bibitem{FerrierPRL2017}
\bibinfo{author}{Ferrier, M.} \emph{et~al.}
\newblock \bibinfo{title}{Quantum fluctuations along symmetry crossover in a
  {Kondo}-correlated quantum dot}.
\newblock \emph{\bibinfo{journal}{Physical Review Letters}}
  \textbf{\bibinfo{volume}{118}}, \bibinfo{pages}{196803}
  (\bibinfo{year}{2017}).

\bibitem{HataArxiv2018}
\bibinfo{author}{Hata, T.} \emph{et~al.}
\newblock \bibinfo{title}{Enhanced shot noise of multiple {Andreev} reflections
  in a carbon nanotube quantum dot in {SU(2)} and {SU(4)} {Kondo} regimes}.
\newblock \emph{\bibinfo{journal}{Physical Review Letters}}
  \textbf{\bibinfo{volume}{121}}, \bibinfo{pages}{247703}
  (\bibinfo{year}{2018}).

\bibitem{Hewson2005}
\bibinfo{author}{Hewson, A.~C.}, \bibinfo{author}{Bauer, J.} \&
  \bibinfo{author}{Oguri, A.}
\newblock \bibinfo{title}{Non-equilibrium differential conductance through a
  quantum dot in a magnetic field}.
\newblock \emph{\bibinfo{journal}{Journal of Physics: Condensed Matter}}
  \textbf{\bibinfo{volume}{17}}, \bibinfo{pages}{5413} (\bibinfo{year}{2005}).

\bibitem{TakadaPRL2014}
\bibinfo{author}{Takada, S.} \emph{et~al.}
\newblock \bibinfo{title}{Transmission phase in the {Kondo} regime revealed in
  a two-path interferometer}.
\newblock \emph{\bibinfo{journal}{Physical Review Letters}}
  \textbf{\bibinfo{volume}{113}}, \bibinfo{pages}{126601}
  (\bibinfo{year}{2014}).

\bibitem{TakadaPRB2016}
\bibinfo{author}{Takada, S.} \emph{et~al.}
\newblock \bibinfo{title}{Low-temperature behavior of transmission phase shift
  across a {Kondo} correlated quantum dot}.
\newblock \emph{\bibinfo{journal}{Physical Review B}}
  \textbf{\bibinfo{volume}{94}}, \bibinfo{pages}{081303}
  (\bibinfo{year}{2016}).

\bibitem{DelagPhD}
\bibinfo{author}{Delagrange, R.}
\newblock \emph{\bibinfo{title}{Josephson effect and high frequency emission in
  a carbon nanotube in the Kondo regime}}.
\newblock Ph.D. thesis, \bibinfo{school}{Universit{\'e} Paris Saclay}
  (\bibinfo{year}{2016}).


\bibitem{vanderWielPRL2002}
\bibinfo{author}{van~der Wiel, W.~G.} \emph{et~al.}
\newblock \bibinfo{title}{Two-stage {Kondo} effect in a quantum dot at a high
  magnetic field}.
\newblock \emph{\bibinfo{journal}{Physical Review Letters}}
  \textbf{\bibinfo{volume}{88}}, \bibinfo{pages}{126803}
  (\bibinfo{year}{2002}).

\bibitem{Bulla2008}
\bibinfo{author}{Bulla, R.}, \bibinfo{author}{Costi, T.~A.} \&
  \bibinfo{author}{Pruschke, T.}
\newblock \bibinfo{title}{Numerical renormalization group method for quantum
  impurity systems}.
\newblock \emph{\bibinfo{journal}{Reviews of Modern Physics}}
  \textbf{\bibinfo{volume}{80}}, \bibinfo{pages}{395--450}
  (\bibinfo{year}{2008}).


\end{thebibliography}
\end{document}



\title{\mdseries{Supplementary Information for}

\it{Three-body correlations in nonlinear response of correlated quantum liquid}}

\author{Tokuro Hata\footnote{email: hata@phys.titech.ac.jp
}}
\affiliation{Graduate School of Science, Osaka University, Toyonaka, Osaka 560-0043, Japan.}

\author{Yoshimichi Teratani}
\affiliation{Department of Physics, Osaka City University, Osaka 558-8585, Japan.}

\author{Tomonori Arakawa}
\affiliation{Graduate School of Science, Osaka University, Toyonaka, Osaka 560-0043, Japan.}
\affiliation{Center for Spintronics Research Network, Osaka University, Toyonaka, Osaka 560-8531, Japan.}

\author{Sanghyun Lee}
\affiliation{Graduate School of Science, Osaka University, Toyonaka, Osaka 560-0043, Japan.}

\author{Meydi Ferrier}
\affiliation{Graduate School of Science, Osaka University, Toyonaka, Osaka 560-0043, Japan.}
\affiliation{Université Paris-Saclay, CNRS, Laboratoire de Physique des Solides, 91405, Orsay, France.}

\author{Richard Deblock}
\affiliation{Université Paris-Saclay, CNRS, Laboratoire de Physique des Solides, 91405, Orsay, France.}

\author{Rui Sakano}
\affiliation{The Institute for Solid State Physics, The University of Tokyo, Chiba 277-8581, Japan.}

\author{Akira Oguri}
\affiliation{Department of Physics, Osaka City University, Osaka 558-8585, Japan.}
\affiliation{Nambu Yoichiro Institute of Theoretical and Experimental Physics, Osaka City University, Osaka 558-8585, Japan.}

\author{Kensuke Kobayashi\footnote{email: kensuke@phys.s.u-tokyo.ac.jp
}}
\affiliation{Graduate School of Science, Osaka University, Toyonaka, Osaka 560-0043, Japan.}
\affiliation{Institute for Physics of Intelligence and Department of Physics, The University of Tokyo, Bunkyo-ku, Tokyo 113-0033, Japan.}
\affiliation{Trans-scale Quantum Science Institute, The University of Tokyo, Bunkyo-ku, Tokyo 113-0033, Japan.}

\maketitle
\section*{SUPPLEMENTARY NOTE 1}
\subsection*{\label{sec:001}Two- and three body correlations in correlated quantum liquid~\cite{OguriPRL2018}}
Let us consider a quantum dot (QD) with a single electron energy $\varepsilon_\sigma=\varepsilon-\sigma\mu_{\rm B}B$, where $\sigma$ is a spin $\sigma~(=\uparrow, \downarrow {\rm or}\ +, -)$.
$\mu_{\rm B}$ is the Bohr magneton. The $g$-factor of electrons is taken as 2.
$\varepsilon=-U/2$ and $B=0$ when the QD holds time-reversal symmetry (TRS) and particle-hole symmetry (PHS) [see the center of Fig. 1c in the main text].
For the Hamiltonian ($\cal H$) of this system, the occupation number of the impurity state $\langle n_\sigma \rangle$ at temperature $T$ is given using the free energy $\Omega \equiv -k_{\rm B} T \ln [\textrm{Tr} e^{-{\cal H}/k_{\rm B} T}]$ such that $\langle n_\sigma \rangle = \partial \Omega / \partial \varepsilon_\sigma$ ($k_{\rm B}$ is the Boltzmann constant). The Friedel sum rule relates the phase shift $\delta_\sigma$ to the occupation number; $\langle n_\sigma \rangle \rightarrow \delta_\sigma /\pi$ at $T \rightarrow 0$.

Now, following Ref.~\cite{OguriPRL2018}, we define the susceptibility as:
\begin{equation}
\chi_{\sigma_1 \sigma_2}\equiv -\dfrac{\partial^2 \Omega}{\partial \varepsilon_{\sigma_1} \partial \varepsilon_{\sigma_2}} = -\dfrac{\partial \langle n_{\sigma_1} \rangle}{\partial \varepsilon_{\sigma_2}}.
\end{equation}
The Kondo temperature and the Wilson ratio are described by the susceptibility: $T^{\ast}_{\rm K} \equiv 1/(4\sqrt{\chi_{\uparrow\uparrow}\chi_{\downarrow\downarrow}})$ and $R\equiv 1-\chi_{\uparrow\downarrow}/\sqrt{\chi_{\uparrow\uparrow}\chi_{\downarrow\downarrow}}$~\cite{NozieresJLTP1974, YosidaPTPS1970}.
Note that $k_{\rm B}T^{\ast}_{\rm K} = 1/(4\chi_{\uparrow\uparrow})$ and $R= 1-\chi_{\uparrow\downarrow}/\chi_{\uparrow\uparrow}$ when either $\epsilon=-U/2$ or $B=0$ because of $\chi_{\uparrow\uparrow}=\chi_{\downarrow\downarrow}$.
The conventional charge and spin susceptibilities are given by the linear combinations of $\chi_{\sigma_1 \sigma_2}$; $\chi_C=\chi_{\uparrow \uparrow}+\chi_{\downarrow \downarrow}+\chi_{\uparrow \downarrow}+\chi_{\downarrow \uparrow}$ and $\chi_s=\frac{1}{4}(\chi_{\uparrow \uparrow}+\chi_{\downarrow \downarrow}-\chi_{\uparrow \downarrow}-\chi_{\downarrow \uparrow})$, respectively.
Thus, $\chi_{\sigma_1 \sigma_2}$ contains sufficient information to describe the system in the linear response regime.
The susceptibility can also be expressed as:
\begin{equation}
\chi_{\sigma_1 \sigma_2}=\int_{0}^{1/T} d\tau \left<\delta n_{\sigma_1}(\tau)\delta n_{\sigma_2}\right>,
\end{equation}
where $\delta n_{\sigma}\equiv n_{\sigma}-\left<n_{\sigma}\right>$.

The nonlinear susceptibility, which is the third derivative of the free energy, contributes to the next leading Fermi-liquid corrections when the energy level is away from half filling (see the right and left of Fig. 1c):
\begin{equation}
\chi_{\sigma_1 \sigma_2 \sigma_3} \equiv - \frac{\partial^3 \Omega}{\partial \varepsilon_{\sigma_1}\partial \varepsilon_{\sigma_2}\partial \varepsilon_{\sigma_3}}=\frac{\partial \chi_{\sigma_2 \sigma_3}}{\partial \varepsilon_{\sigma_1}}.
\end{equation}
It can also be expressed as:
\begin{equation}
\chi_{\sigma_1 \sigma_2 \sigma_3} =-\int_{0}^{1/T}d\tau_3\int_{0}^{1/T}d\tau_2\left<T_{\tau}\delta n_{\sigma_3}(\tau_3)\delta n_{\sigma_2}(\tau_2)\delta n_{\sigma_1}\right>,
\end{equation}
where  $T_{\tau}$ denotes the time-ordering product.

Now, we consider the Kondo-correlated QD when either TRS or PHS is broken by using the following expression of differential conductance:
\begin{equation}
\dfrac{dI}{dV}=\dfrac{e^2}{h}\left(\sum_{\sigma}\sin^2{\delta_\sigma}\right)-{\alpha_{\rm V}}\left(\dfrac{eV}{k_{\rm B}T_{\rm K}}\right)^2-{\alpha_{\rm T}}\left(\dfrac{\pi T}{T_{\rm K}}\right)^2 +\cdots.
\end{equation}
Here, $T_{\rm K}$ is the Kondo temperature at the TRS and the PHS point: $T_{\rm K}^{\ast}(\varepsilon/U=-0.5, B=0\ {\rm T})$.
The phase shift is defined as:
\begin{equation}
\cot{\delta_{\sigma}\equiv}\dfrac{\varepsilon_{\sigma}+\Sigma_\sigma}{\gamma}, 
\label{eqn06}
\end{equation}
where $\Sigma_\sigma$ is the self-energy at $eV=0$ and zero-frequency, and $\gamma$ is the half width of energy levels, $\gamma=\Gamma/2$, in our definition.

$\alpha_{\rm V}$ consists of $\chi_{\sigma_1\sigma_2}$ and $\chi_{\sigma_1\sigma_2\sigma_3}$ as follows:
\begin{equation}
\begin{split}
\alpha_{\rm V}&=\dfrac{2e^2}{h}\dfrac{(k_{\rm B}T_{\rm K})^2}{2}\sum_{\sigma}\dfrac{\pi^2}{4}\left[-\cos{2\delta_{\sigma}}\left(\chi^{2}_{\sigma\sigma}+5\chi^{2}_{\sigma-\sigma}\right)+\dfrac{\sin{2\delta_{\sigma}}}{2\pi}\left(\chi_{\sigma\sigma\sigma}+3\chi_{\sigma-\sigma-\sigma}\right)\right]\\
&=\dfrac{2e^2}{h}\dfrac{\pi^2}{64}\left(\dfrac{T_{\rm K}}{T_{\rm K}^{\ast}}\right)^2\dfrac{1}{2\chi_{\uparrow\uparrow}^2}\sum_{\sigma}\left[-\cos{2\delta_{\sigma}}\times\left(\chi^{2}_{\sigma\sigma}+5\chi^{2}_{\sigma-\sigma}\right)+\dfrac{\sin{2\delta_{\sigma}}}{2\pi}\left(\chi_{\sigma\sigma\sigma}+3\chi_{\sigma-\sigma-\sigma}\right)\right]\\
&\equiv\dfrac{2e^2}{h}\dfrac{\pi^2}{64}\left(W_2+W_3\right)\times\left(\dfrac{T_{\rm K}}{T_{\rm K}^{\ast}}\right)^2, \label{eqn6}
\end{split}
\end{equation}
where we use $k_{\rm B}T_{\rm K}^{\ast}=1/4\chi_{\uparrow\uparrow}$.
We define $W_2$ and $W_3$ as two- and three-body correlations, respectively, which we evaluate in the main text. 

While it is not treated in this work, it is noted that the coefficient of $(\pi T/T_{\rm K})^2$, $\alpha_{\rm T}$, is: 
\begin{equation}
\begin{split}
\alpha_{\rm T}&=\dfrac{2e^2}{h}\dfrac{\pi^2}{48}\left(\dfrac{T_{\rm K}}{T_{\rm K}^{\ast}}\right)^2\dfrac{1}{2\chi_{\uparrow\uparrow}^2}\sum_{\sigma}\left[-\cos{2\delta_{\sigma}}\times\left(\chi^{2}_{\sigma\sigma}+2\chi^{2}_{\sigma-\sigma}\right)+\dfrac{\sin{2\delta_{\sigma}}}{2\pi}\left(\chi_{\sigma\sigma\sigma}+\chi_{\sigma-\sigma-\sigma}\right)\right].
\end{split}
\end{equation}

\subsection*{\label{sec:002}The free particle (FP) model}
\subsubsection*{General formulas}
In the main text, we compare the experimental result with the $U=0$ case, which we refer as the free particle (FP) model. We explain the detail of the FP model here.

The current for non-interacting electrons ($U=0$) at zero temperature is:
\begin{align}
I=\dfrac{e}{h}\int_{-eV/2}^{eV/2}d\omega\sum_{\sigma}\mathcal{T}_{\sigma}(\omega),
\label{eqn001}
\end{align}
where $\mathcal{T}_{\sigma}(\omega)$ is transmission at frequency, $\omega$:
\begin{align}
\mathcal{T}_{\sigma}(\omega)=\dfrac{\gamma_0^2}{(\omega-\varepsilon_{\sigma})^2+\gamma_0^2}.
\label{eqn002}
\end{align}
$2\gamma_0$ corresponds to the half width of a resonance peak.
$\varepsilon_{\sigma}$ is a single electron energy, where $\varepsilon_{\sigma}=\varepsilon-\sigma\mu_{\rm B}B$ with spin, $\sigma$. 
The $g$-factor of electrons is taken as 2.
The differential conductance ($dI/dV$) is expressed as:
\begin{align}
\dfrac{dI}{dV}&=\dfrac{e^2}{h}\sum_{\sigma}\dfrac{1}{2}\left[\mathcal{T}_{\sigma}(eV/2)+\mathcal{T}_{\sigma}(-eV/2)\right]\label{eqn003}\\
&=\dfrac{e^2}{h}\sum_{\sigma}\dfrac{1}{2}\left[\dfrac{\gamma_0^2}{(eV/2-\varepsilon_{\sigma})^2+\gamma_0^2}+\dfrac{\gamma_0^2}{(eV/2+\varepsilon_{\sigma})^2+\gamma_0^2}\right]\label{eqn004}\\ 
&=\dfrac{e^2}{h}\left[\sum_{\sigma}\dfrac{1}{1+(\varepsilon_{\sigma}/\gamma_0)^2}\right]-\alpha_{\rm V}^{(0)}\left(\dfrac{eV}{2\gamma_0}\right)^2+\mathrm{O}(V^4),
\label{eqn005}
\end{align}
where $\alpha_{\rm V}^{(0)}$ in the last line of the above equation is:
\begin{align}
\alpha_{\rm V}^{(0)}\equiv \dfrac{e^2}{h}\sum_{\sigma}\dfrac{1-3(\varepsilon_{\sigma}/\gamma_0)^2}{\left\{1+(\varepsilon_{\sigma}/\gamma_0)^2\right\}^3}.\label{eqn006}
\end{align}
Because phase shift $\delta_\sigma^{(0)}$ at $U=0$ is given by $\cot{\delta_{\sigma}^{(0)}}=\varepsilon_\sigma/\gamma_0$ [see also Supplementary Equation (\ref{eqn06})], the first term of Supplementary Equation (\ref{eqn005}) can be replaced by $\sin^2{\delta_{\sigma}^{(0)}}$:
\begin{align}
\sin^2{\delta_{\sigma}^{(0)}}=\dfrac{1}{1+(\varepsilon_\sigma/\gamma_0)^2}.\label{eqn007}
\end{align}

When either the TRS or PHS is preserved, $\sin^2{\delta_{\uparrow}^{(0)}}=\sin^2{\delta_{\downarrow}^{(0)}}$, $\varepsilon_{\uparrow}^2=\varepsilon_{\downarrow}^2$, and $\chi_{\uparrow\uparrow}^{(0)}=\chi_{\downarrow\downarrow}^{(0)}$.
In terms of susceptibility, we get $\alpha_{\rm V}^{(0)}$ by replacing $k_{\rm B}T_{\rm K}$ with $2\gamma_0$ in Supplementary Equation (\ref{eqn6}):
\begin{align}
\alpha_{\rm V}^{(0)}&=\dfrac{e^2}{h}\dfrac{\pi^2}{2}\left(W_2^{(0)}+W_3^{(0)}\right)\times(2\gamma_0)^2\times \chi_{\uparrow\uparrow}^{(0)2} \notag \\
&=\dfrac{2e^2}{h}\sin^4{\delta_{\uparrow}}\left(W_2^{(0)}+W_3^{(0)}\right).
\end{align}
$W_{2}^{(0)}$ and $W_{3}^{(0)}$ are defined as follows:
\begin{align}
W_{2}&=-\dfrac{1}{2\chi_{\uparrow\uparrow}^{2}}\sum_{\sigma}\cos{2\delta_{\sigma}}\times\left(\chi^{2}_{\sigma\sigma}+5\chi^{2}_{\sigma-\sigma}\right)\notag \\
&\underset{U=0}{\longrightarrow} -\dfrac{1}{2}\sum_{\sigma}\cos{2\delta^{(0)}_{\sigma}}\ (\because \chi^{(0)}_{\sigma-\sigma}=0) \notag \\
&=\dfrac{1}{2}\sum_{\sigma}\dfrac{\gamma_0^2-\varepsilon_\sigma^2}{\varepsilon_\sigma^2+\gamma_0^2}
\equiv W_{2}^{(0)}.\\
W_{3}&=\dfrac{1}{2\chi_{\uparrow\uparrow}^{2}}\sum_{\sigma}\dfrac{\sin{2\delta_{\sigma}}}{2\pi}\left(\chi_{\sigma\sigma\sigma}+3\chi_{\sigma-\sigma-\sigma}\right) \notag \\
&\underset{U=0}{\longrightarrow} \dfrac{1}{2\chi_{\uparrow\uparrow}^{(0)2}}\sum_{\sigma}\dfrac{\sin{2\delta^{(0)}_{\sigma}}}{2\pi}\chi^{(0)}_{\sigma\sigma\sigma}\ (\because \chi^{(0)}_{\sigma-\sigma-\sigma}=0) \notag \\
&=-\sum_{\sigma}\dfrac{\varepsilon_\sigma^2}{\varepsilon_\sigma^2+\gamma_0^2}\equiv W_{3}^{(0)}.
\end{align}
We use the following relations for the calculations:
\begin{align}
\chi_{\sigma\sigma}^{(0)}&=\dfrac{1}{\pi}\dfrac{\gamma_0}{\varepsilon^2_{\sigma}+\gamma^2_0}=\dfrac{\sin^2{\delta_{\sigma}^{(0)}}}{\pi\gamma_0},\\
\chi_{\sigma\sigma\sigma}^{(0)}&\equiv\dfrac{\partial\chi_{\sigma\sigma}^{(0)}}{\partial\varepsilon_{\sigma}}=\dfrac{1}{\pi}\dfrac{-2\gamma_0\varepsilon_{\sigma}}{(\varepsilon_{\sigma}^2+\gamma_0^2)^2}=-\dfrac{2}{\pi\gamma_0^2}\cot{\delta_{\sigma}^{(0)}}\sin^2{\delta^{(0)}_{\sigma}}\notag \\
&=-2\pi\cot{\delta_{\sigma^{(0)}}}(\chi_{\sigma\sigma}^{(0)})^2
\end{align}

\subsubsection*{Evaluation of $\alpha_{\rm V}$ under magnetic field}
Supplementary Equation (\ref{eqn006}) is written as follows in the TRS breaking regime ($B \neq 0$): 
\begin{align}
\alpha_{\rm V}^{(0)}(B)=\dfrac{2e^2}{h}\dfrac{1-3(\mu_{\rm B}B/\gamma_0)^2}{\left\{1+(\mu_{\rm B}B/\gamma_0)^2\right\}^3},
\label{eqn013}
\end{align}
showing that the magnetic field dependence of $\alpha_{\rm V}^{(0)}$ is determined by the value of $\gamma_0$ at zero field.

\begin{figure}
\center \includegraphics[width=70mm]{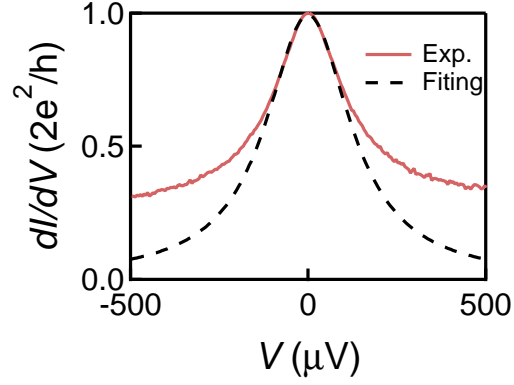}
\caption{The differential conductance ($dI/dV$) as a function of source voltage ($V$) at zero field.}
\label{002}
\end{figure}

We regard the experimental Kondo peak at zero field as if it were a resonance peak with $U=0$ and obtain $\gamma_0$ by fitting the experimental result with the following formula:
\begin{align}
\dfrac{dI}{dV}=\dfrac{2e^2}{h}\dfrac{\gamma_0^2}{(eV/2)^2+\gamma_0^2}.\label{eqn014}
\end{align}
Supplementary Figure \ref{002} shows the fitting result for the source-drain voltage dependence ($V$) of differential conductance ($dI/dV$) at zero field (see the detail of the experimental setup and definition of zero field in the main text).
The evaluated $2\gamma_0$ is $130\ {\rm \mu eV}(=1.5\ {\rm K})$, which is comparable with the Kondo temperature given by the temperature dependence of the zero bias conductance (see the main text and also the next paragraph). 
We show the magnetic field dependence of $\alpha_{\rm V}^{(0)}$ calculated with Supplementary Equation (\ref{eqn013}) and $\gamma_0=65\ {\rm \mu eV}$ in Fig. 3b in the main text.
Similarly, we obtain the magnetic field dependence of $W_2^{(0)}$ and $W_3^{(0)}$ with:
\begin{align}
W_{2}&=\dfrac{\gamma_0^2-(\mu_{\rm B}B)^2}{(\mu_{\rm B}B)^2+\gamma_0^2}\\
W_{3}&=\dfrac{-2(\mu_{\rm B}B)^2}{(\mu_{\rm B}B)^2+\gamma_0^2},
\end{align}
which are shown in Figs. 3a and 3c in the main text.

In our experiment, as $k_{\rm B}T_{\rm K}=138\ {\rm \mu eV}$ and $\gamma_0=65\ {\rm \mu eV}$, $k_{\rm B}T_{\rm K}/\gamma_0\sim2$ is the case. This is expected for $R\rightarrow 2$, which can be explained as follows.
In the TRS and PHS case, the following is established \cite{OguriJPSJ2005},
\begin{align}
\frac{dI}{dV}=\dfrac{2e^2}{h}\left[ 1- \frac{1+5(R-1)^2}{4} \left(\frac{eV}{\tilde{\Gamma}}\right)^2\right].
\end{align}
Here, $\tilde{\Gamma}$ is the renormalized width of the Kondo resonance, which depends on $U$ and $\Gamma$. 
For $U/\Gamma \rightarrow \infty$ ($R\rightarrow 2$),   $\tilde{\Gamma}=4k_{\rm B}T_{\rm K} /\pi$ holds, and thus,
\begin{align}
\frac{dI}{dV}=\frac{2e^2}{h}\left[ 1-\frac{3}{2}\left(\frac{\pi}{4}\right)^2 \left(\frac{eV}{k_{\rm B}T_{\rm K}}\right)^2\right].
\end{align}
On the other hand, for $\vert eV/2\gamma_0 \vert \ll 1$, Equation (\ref{eqn014}) yields
\begin{align}
\frac{dI}{dV}=\dfrac{2e^2}{h}\dfrac{\gamma_0^2}{(eV/2)^2+\gamma_0^2} \rightarrow \frac{2e^2}{h}\left[ 1- \left(\frac{eV}{2\gamma_0}\right)^2\right].
\end{align}
Comparing these two, the following is obtained.
\begin{align}
\frac{k_{\rm B}T_{\rm K}}{\gamma_0}=\sqrt{\frac{3}{8}}\pi = 1.923\cdots.
\end{align}
It is interesting to note that, when $T_{\rm K}$ is given, the differential conductance tells us information on $R$ in this way.

\clearpage

\section*{SUPPLEMENTARY NOTE 2}
\subsection*{The analysis in the Kondo regime}
\subsubsection*{Magnetic field dependence of Kondo temperature}
The solid curves in Supplementary Figure \ref{003}(a) are experimental conductance at different magnetic fields.
We obtain $\gamma_0(B)$ and $G_0(B)$ for each field by fitting the experimental results with Supplementary Equation (\ref{eqn004}).
Supplementary Figure \ref{003}(b) shows $\mu_{\rm B}B/k_{\rm B}T_{\rm K}$ or $B$ dependence of $\gamma_0(0)/\gamma_0(B)$ and $G_0(B)$.
$T_{\rm K}$ is the Kondo temperature at zero field, which we evaluated from the temperature dependence (see the main text).

In the $U=0$ case, $T^{\ast}_{\rm K}(B)$, $\gamma_0(B)$, and phase shift $\delta_\sigma$ are related as:
\begin{align}
4k_{\rm B}T^{\ast}_{\rm K}=\dfrac{\pi\gamma_0(B)}{\sin^2{\delta_\sigma}}.
\label{eqn016}
\end{align}
Here, $T^{\ast}_{\rm K}$ is a characteristic temperature nominally obtained according to the definition of 
$T^{\ast}_{\rm K} \equiv 1/(4\sqrt{\chi_{\uparrow\uparrow}\chi_{\downarrow\downarrow}})$.

Thus, we obtain the following relation between $T^{\ast}_{\rm K}$ and $\gamma_0(B)$:
\begin{align}
\dfrac{T_{\rm K}^{\ast}(0)}{T_{\rm K}^{\ast}(B)}=\dfrac{\gamma_0(0)}{\gamma_0(B)}\dfrac{G(B)}{G(0)},
\label{eqn017}
\end{align}
Thus, we get $T_{\rm K}^{\ast}(0)/T_{\rm K}^{\ast}(B)$ by multiplying $\gamma_0(0)/\gamma_0(B)$ and $G_0(B)$ [the red circles in Supplementary Figure \ref{003}(c).
The evaluation agrees well with the theoretical curve given by NRG calculations with $U/\Gamma=3.5$~\cite{OguriPRB2018_2}.

\begin{figure}[h]
\center \includegraphics[width=150mm]{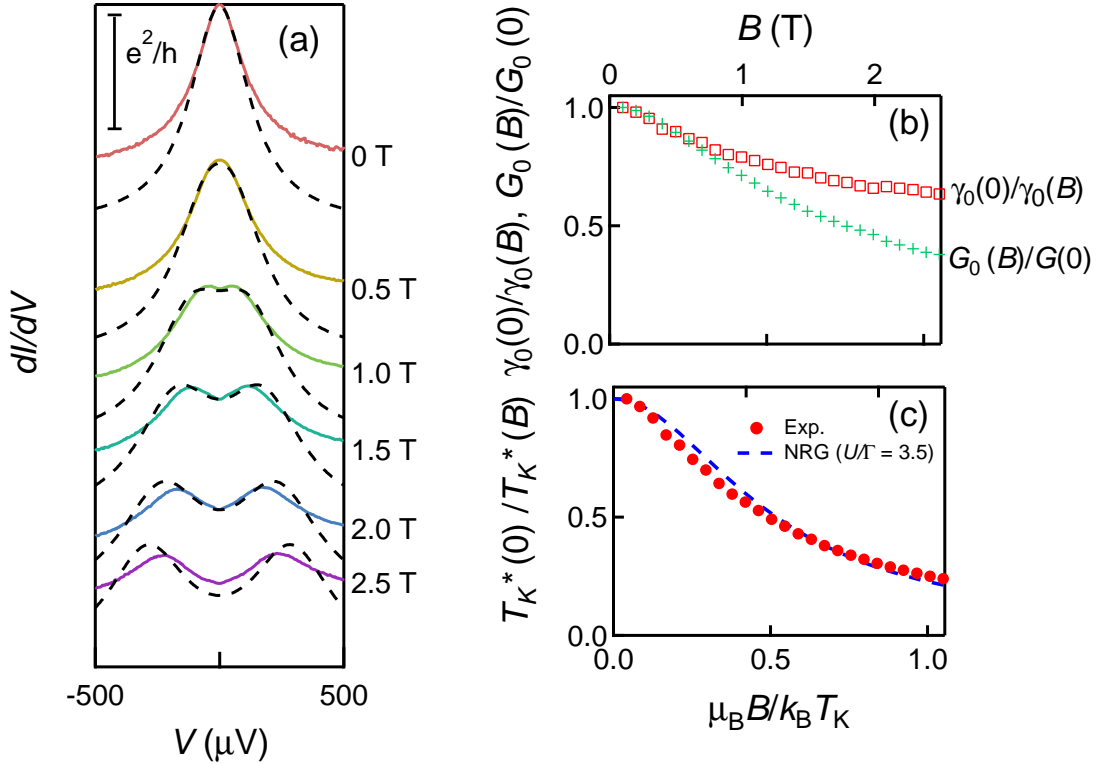}
\caption{(a) $dI/dV$ as a function of $V$ for different magnetic fields. The dashed lines are fitting curves with Supplementary Equation (\ref{eqn004}). (b) The red squares are $\gamma_0(0)/\gamma_0(B)$ obtained from the fitting. The green cross marks are the zero bias conductance. (c) The red circles are $T_{\rm K}^{\ast}(0)/T_{\rm K}^{\ast}(B)$, which we obtained by multiplying $\gamma_0(0)/\gamma_0(B)$ with $G(B)/G(0)$. The dashed line is given by the NRG calculations with $U/\Gamma=3.5$.}
\label{003}
\end{figure}

\subsubsection*{Small dip in conductance at finite magnetic field.}

\begin{figure}[!h]
\center \includegraphics[width=120mm]{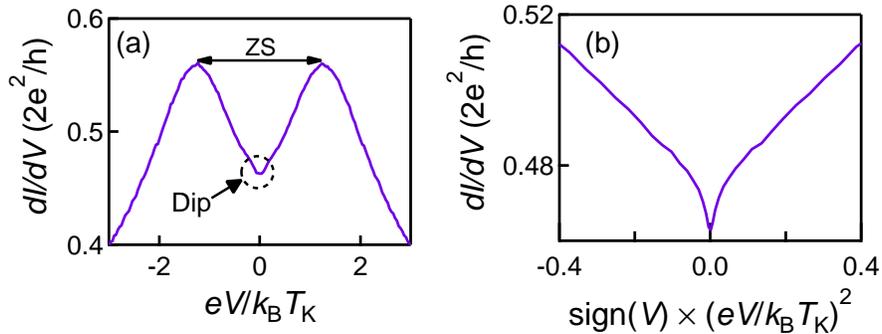}
\caption{(a) $dI/dV$ as a function of $eV/k_{\rm B}T_{\rm K}$ at $2\ {\rm T}$. The circle points out a very small dip or cusp-like structure just around the zero bias, which might be attributed to the two-stage Kondo effect. (b) $dI/dV$ as a function of ${\rm sign}(V)\times(eV/k_{\rm B}T_{\rm K})^2$ at $2\ {\rm T}$. The small dip in the left figure is attributed to the sharp slope around zero bias.}
\label{004}
\end{figure}

Supplementary Figure \ref{004}(a) is $dI/dV$ as a function of $eV/k_{\rm B}T_{\rm K}$ at $2\ \rm{T}$ ($\mu_{\rm B}B/k_{\rm B}T_{\rm K}=0.84$), where Zeeman splitting (ZS) is clearly seen.
There is a small dip or cusp-like structure just around the zero bias, whose shape is not parabolic to cause a sharp change in $dI/dV$ as a function of ${\rm sign}(V)\times(eV/k_{\rm B}T_{\rm K})^2$ [Supplementary Figure \ref{004}(b)].
This may be attributed to the two-stage Kondo effect~\cite{vanderWielPRL2002} and we do not use several points near zero bias to avoid this effect.

\clearpage

\subsubsection*{Magnetic field and gate voltage dependence of the Wilson ratio}
\begin{figure}[!h]
\center \includegraphics[width=80mm]{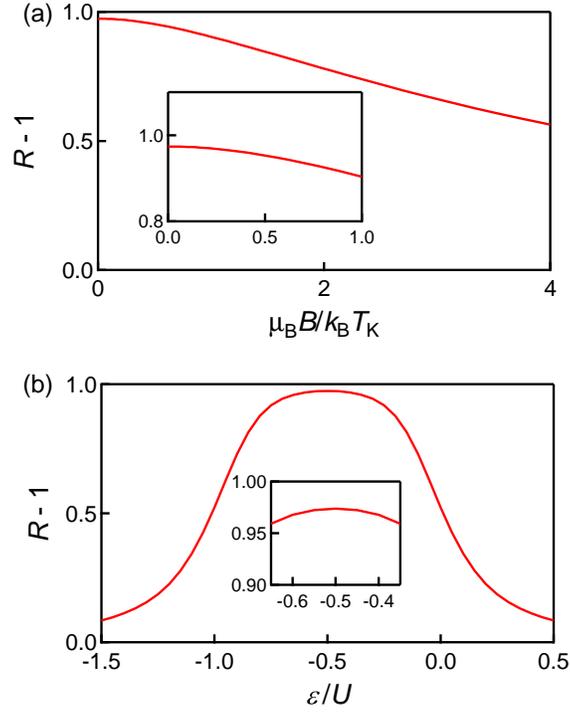}
\caption{(a) (b) $R-1$ as a function of $\mu_{\rm B}B/k_{\rm B}T_{\rm K}$ and $\varepsilon/U$, respectively, which are given by the NRG calculations with $U/\Gamma=3.5$~\cite{OguriPRB2018_2}.}
\label{005}
\end{figure}

Supplementary Figures \ref{005} (a) and (b) are $\mu_{\rm B}B/k_{\rm B}T_{\rm K}$ and $\varepsilon/U$ dependence of $R-1$, which are given by the NRG calculations with $U/\Gamma=3.5$, respectively~\cite{OguriPRB2018_2}.
Here, $R$ is the Wilson ratio.
The Wilson Ratio decreases as the time-reversal and particle-hole symmetry are broken.
The insets are the expanded views of $R-1$ whose horizontal scale corresponds to those shown in the main text, showing the value does not change in these regions.
Thus, we use experimental value of $R=1.95$ at zero field and at $\varepsilon=-U/2$ for our analysis.

\clearpage

\section*{SUPPLEMENTARY NOTE 3}
\subsection*{Analysis procedure}
We show the analysis procedures to obtain $W_2$ and $W_3$ as a function of magnetic field and gate voltage as follows.
\subsubsection*{Magnetic field dependence of $W_2$}
	\begin{enumerate}
	\item We obtain $T_{\rm K}$ by analyzing the temperature dependence of $G_{0}$~\cite{FerrierNatPhys2016}.
	\item We obtain $T_{\rm K}^{\ast}(\varepsilon=0,\ B)$ by analyzing $dI/dV$ at each field [Supplementary Note 2 and Supplementary Fig. 2(c)].
	\item We derive $m_d(B)$ from $G_0(B)$ by using $G_0=\dfrac{2e^2}{h}\cos^2{(\pi m_d)}$ (Fig. 3a in the main text).
	\item We obtain $R=1.95$ at $B=0\ {\rm T}$ by the shot noise measurement~\cite{FerrierNatPhys2016}.
	\item We assume that $R$ is constant up to $\mu_{B}B/k_{\rm B}T_{\rm K}\sim1$, which is justified by the NRG calculation [Supplementary Note 2 and Supplementary Fig. 4(a)].
 	\item We obtain $W_2$ as a function of $\mu_{\rm B}B/k_{\rm B}T_{\rm K}$ by using $W_2=-\left[1+5(R-1)^2\right]\cos{(\pi m_d)}$ (Fig. 3a in the main text).
	\end{enumerate}

\subsubsection*{Magnetic field dependence of $W_3$}
	\begin{enumerate}
	\item We plot $dI/dV$ as a function of ${\rm sign}(V)\times(eV/k_{\rm B}T_{\rm K})^2$, and obtain $\alpha_{\rm V}$ at each magnetic field by fitting the plotted points with $dI/dV=G_0-\alpha_{\rm V}(eV/k_{\rm B} T_{\rm K})^2$ (Fig. 3b in the main text).
	\item We calculate $W_3$ by using $\alpha_{\rm V}$, $W_2$, $T_{\rm K}$, and $T_{\rm K}^{\ast}$ with Equation (6) in the main text (Fig. 3c in the main text).
   \end{enumerate}

\subsubsection*{Gate voltage dependence of $W_2$}
	\begin{enumerate}
	\item We obtain $T_{\rm K}^{\ast}(\varepsilon,\ B=0)$ by analyzing the temperature dependence of $G_0$ at each gate voltage [Fig. 2a in the main text].
	\item We derive $\delta_{\sigma}(\varepsilon)$ from $G_0(\varepsilon)$ by using $G_0=\dfrac{2e^2}{h}\sin^2{\delta_{\sigma}}$. Here, we assume that $\delta_{\sigma}=\delta_{\uparrow}=\delta_{\downarrow}$ and that $\delta_{\sigma}>\pi/2$ and  $\delta_{\sigma}<\pi/2$ for $\varepsilon/U<-0.5$ and $\varepsilon/U>-0.5$, respectively [Fig. 4a in the main text].
	\item We assume that $R$ is constant  in a wide region $|\varepsilon/U+0.5|<0.15$, which is justified by the NRG calculation [Section C3 and Supplementary Figure 4(b)].
 	\item We obtain $W_2$ as a function of $\varepsilon/U$ by using $W_2=\left[1+5(R-1)^2\right]\cos{2\delta_{\sigma}}$ [Supplementary Figure 5].
	\end{enumerate}
	
\subsubsection*{Gate voltage dependence of $W_3$}
	\begin{enumerate}
	\item We plot $dI/dV$ as a function of ${\rm sign}(V)\times(eV/k_{\rm B}T_{\rm K})^2$, and obtain $\alpha_{\rm V}$ at each gate voltage by fitting the plotted points with $dI/dV=G_0-\alpha_{\rm V}(eV/k_{\rm B} T_{\rm K})^2$ [Fig. 4b in the main text].
	\item We obtain $W_3$ by using $W_2$, $T_{\rm K}$, $T_{\rm K}^{\ast}$, and Eq. (3) shown in the main text [Fig. 4c in the main text].
   \end{enumerate}

\begin{figure}[htp]
\center \includegraphics[width=80mm]{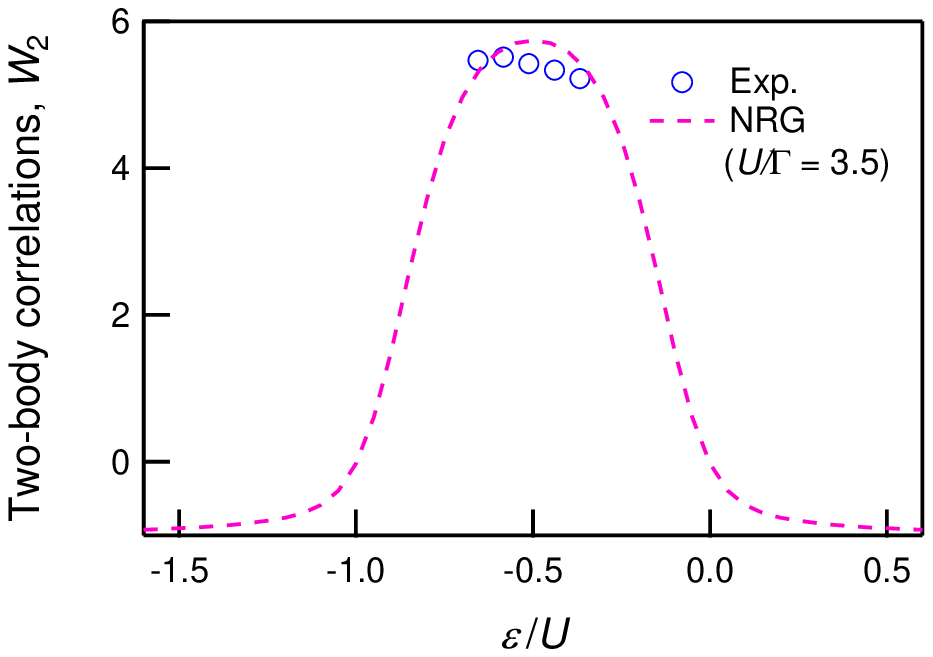}
\caption{$W_2$ as a function of $\varepsilon/U$. The points are experimental results, while the lines are theoretical ones.}
\label{006}
\end{figure}

\clearpage
\section*{SUPPLEMENTARY NOTE 4}
\subsection*{Supplemental data}
\begin{figure}[h]
\center \includegraphics[width=150mm]{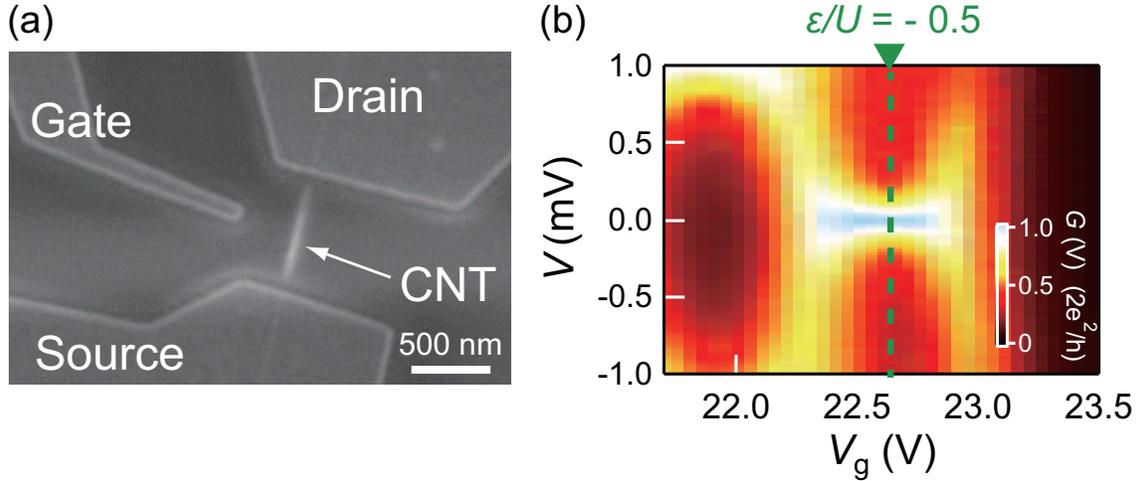}
\caption{(a) Scanning electron micrograph of a carbon nanotube quantum dot. (b) Color plot of $dI/dV$ as a function of $V_{\rm}$ and $V_{\rm g}$. }
\label{SDiamond}
\end{figure}

In this study, we measure the Kondo ridge at $N=3$ in the sample that has been used in Refs.~\cite{FerrierNatPhys2016,FerrierPRL2017,Hata2018}, where $N$ is the number of electrons in the last shell.
Supplementary Figure \ref{SDiamond}(b) is the color plot of $dI/dV$ as a function of $V$ and $V_{\rm g}$, where $V$ and $V_{\rm g}$ are source-drain voltage and gate voltage, respectively.
Here, the magnetic field, $B=0.08\ {\rm T}$, is applied to suppress the superconductivity of the electrodes.
The particle-hole symmetry point, $\varepsilon/U=-0.5$, is also indicated.

We also show $W_2$, $m_d$, $\alpha_{\rm V}$, and $W_3$ as a function of $B$ in Supplementary Figures \ref{Supple8}(a)--(c).
While the same data as a function of the normalized magnetic field $\mu_{\rm B}B/k_{\rm B}T_{\rm K}$ are shown in Figs. 3a, 3b, and 3c in the main text, these graphs shown as a function of $B$ itself might be useful for future analysis.

\begin{figure}[htbp]
\includegraphics[width=100mm]{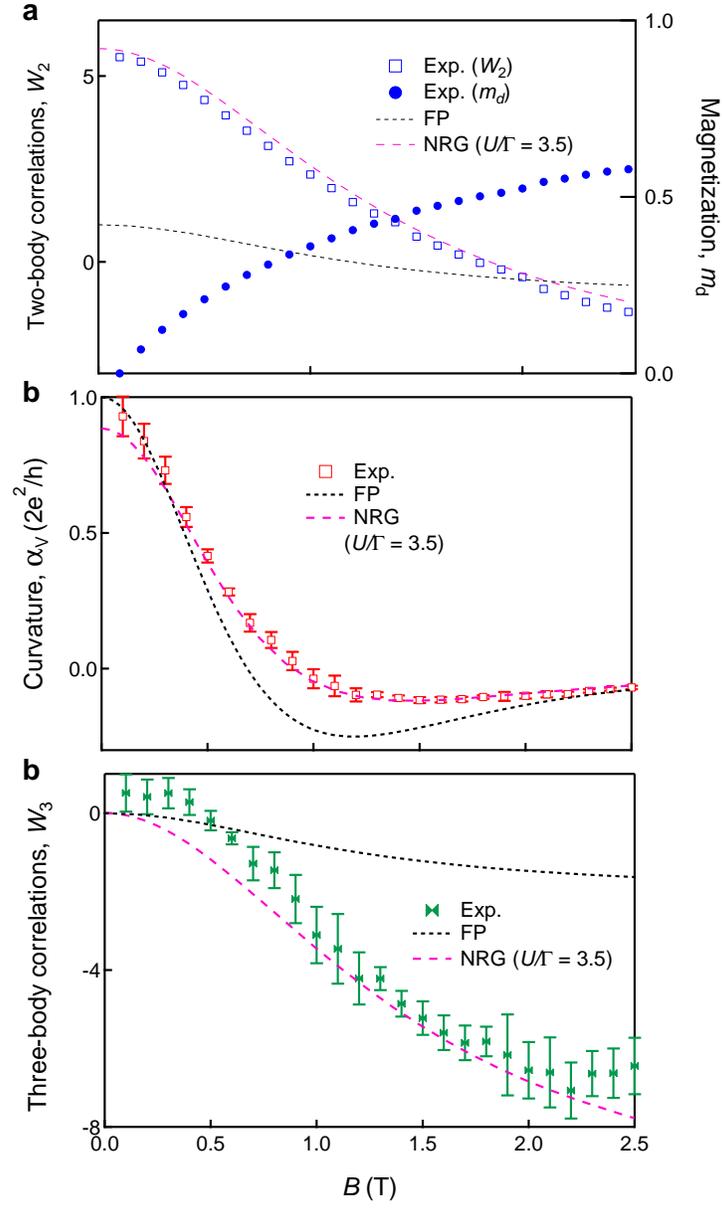}
\caption{(a) -- (c) $W_2$, $m_d$, $\alpha_{\rm V}$, and $W_3$ as a function of magnetic field. The points are experimental results, while the lines are theoretical ones. The dotted and dashed lines for $W_2$ are given by the free particle (FP) model and the NRG calculations ($U/\Gamma=3.5$), respectively.  Error bars in Supplementary Figure 7(b) correspond to the uncertainty of the linear fit performed on slightly different ranges. The error bars in Supplementary Figure 7(c) are determined based on
those of $\alpha_{\rm V}$ shown in Supplementary Figure 7(b).}
\label{Supple8}
\end{figure}

\clearpage
\bibliography{ThreeBodySuppl_ref.bib}